\pdfoutput=1
\documentclass[11pt,letterpaper]{scrartcl}

\makeatletter
\DeclareOldFontCommand{\rm}{\normalfont\rmfamily}{\mathrm}
\DeclareOldFontCommand{\sf}{\normalfont\sffamily}{\mathsf}
\DeclareOldFontCommand{\tt}{\normalfont\ttfamily}{\mathtt}
\DeclareOldFontCommand{\bf}{\normalfont\bfseries}{\mathbf}
\DeclareOldFontCommand{\it}{\normalfont\itshape}{\mathit}
\DeclareOldFontCommand{\sl}{\normalfont\slshape}{\@nomath\sl}
\DeclareOldFontCommand{\sc}{\normalfont\scshape}{\@nomath\sc}
\makeatother

\usepackage[top=2.5cm, bottom=2.5cm, left=3.2cm, right=3cm]{geometry}

\usepackage[T1]{fontenc}
\usepackage[utf8]{inputenc}

\usepackage[hidelinks]{hyperref}
\usepackage{siunitx}
\usepackage[numbers,sort&compress]{natbib}
\usepackage{microtype}
\usepackage{csquotes}
\usepackage{amsmath}
\usepackage{amssymb}
\usepackage{braket}

\usepackage{xcolor}

\usepackage{cleveref}

\usepackage[caption = false]{subfig}
\usepackage{graphicx}

\usepackage{fancyhdr}
\usepackage[auth-sc,affil-it]{authblk}

\usepackage{scalefnt}
\newcommand{\abbrev}{\scalefont{.9}}

\newcommand{\NNLO}{\text{\abbrev NNLO}}
\newcommand{\NLO}{\text{\abbrev NLO}}
\newcommand{\LO}{\text{\abbrev LO}}
\newcommand{\EFT}{\text{\abbrev EFT}}
\newcommand{\SMEFT}{\text{\abbrev SMEFT}}

\newcommand{\SM}{\text{\abbrev SM}}
\newcommand{\BSM}{\text{\abbrev BSM}}
\newcommand{\IR}{\text{\abbrev IR}}
\newcommand{\SU}{\text{\abbrev SU}}
\newcommand{\UV}{\text{\abbrev UV}}
\newcommand{\QCD}{\text{\abbrev QCD}}

\newcommand{\PDF}{\text{\abbrev PDF}}

\newcommand{\LHC}{\text{\abbrev LHC}}
\newcommand{\CMS}{\text{\abbrev CMS}}
\newcommand{\ATLAS}{\text{\abbrev ATLAS}}
\newcommand{\CERN}{\text{\abbrev CERN}}
\newcommand{\CTFOURTEEN}{\text{\abbrev CT14}}
\newcommand{\DDIS}{\text{\abbrev DDIS}}

\newcommand{\CKM}{\text{\abbrev CKM}}

\newcommand{\abs}[1]{\lvert#1\rvert}

\newcommand{\DIANA}{\text{\abbrev DIANA}}
\newcommand{\SaM}{\text{\abbrev S@M}}

\newcommand{\MCFM}{\text{\abbrev MCFM}}

\newcommand{\QD}{\text{\abbrev QD}}
\newcommand{\LANHEP}{\text{\abbrev LANHEP}}
\newcommand{\FORM}{\text{\abbrev FORM}}
\newcommand{\MCNLO}{\text{\abbrev MC@NLO}}
\newcommand{\PHG}{\text{\abbrev POWHEG-BOX}}
\newcommand{\QGRAF}{\text{\abbrev QGRAF}}
\newcommand{\QCDLoop}{\text{\abbrev QCDLoop}}
\newcommand{\Real}{{\mathfrak{Re}}}
\newcommand{\Imag}{{\mathfrak{Im}}}
\newcommand{\IEEE}{\text{\abbrev IEEE}}

\newcommand{\Qone}{\ensuremath{\mathcal{Q}_{\varphi q}^{(3,33)}}}
\newcommand{\Qtwo}{\ensuremath{\mathcal{Q}_{\varphi u d }^{33}}}
\newcommand{\Qthree}{\ensuremath{\mathcal{Q}_{uW}^{33}}}
\newcommand{\Qfour}{\ensuremath{\mathcal{Q}_{dW}^{33}}}
\newcommand{\Qsix}{\ensuremath{\mathcal{Q}_{uG}^{33} }}
\newcommand{\Qseven}{\ensuremath{\mathcal{Q}_{dG}^{33}}}
\newcommand{\Qeight}{\ensuremath{\mathcal{Q}_{4L}}}
\newcommand{\Qnine}{\ensuremath{\mathcal{Q}_{4R}}}
\newcommand{\Qten}{\ensuremath{\mathcal{Q}_{dB}^{33}}}

\newcommand{\Cone}{\ensuremath{\mathcal{C}_{\varphi q}^{(3,33)}}}
\newcommand{\Ctwo}{\ensuremath{\mathcal{C}_{\varphi u d }^{33}}}
\newcommand{\Cthree}{\ensuremath{\mathcal{C}_{uW}^{33}}}
\newcommand{\Cfour}{\ensuremath{\mathcal{C}_{dW}^{33}}}
\newcommand{\Csix}{\ensuremath{\mathcal{C}_{uG}^{33} }}

\newcommand{\Ceight}{\ensuremath{\mathcal{C}_{4L}}}

\newcommand{\cosxprod}{\ensuremath{\cos \theta_{l,x}}}
\newcommand{\cosyprod}{\ensuremath{\cos \theta_{l,y}}}
\newcommand{\coszprod}{\ensuremath{\cos \theta_{l,z}}}

\newcommand{\coslstar}{\ensuremath{\cos \theta_{l}^*}}
\newcommand{\coslN}{\ensuremath{\cos \theta_{l}^N}}
\newcommand{\coslT}{\ensuremath{\cos \theta_{l}^T}}

\def\spa#1.#2{\langle#1\,#2\rangle}
\def\spb#1.#2{[#1\,#2]}

\def\spab#1.#2.#3{\langle\mskip-1mu{#1}
	| #2 | {#3}]}

\def\spba#1.#2.#3{[\mskip-1mu{#1}
	| #2 | {#3}\rangle}

\def\spbb#1.#2.#3.#4{[\mskip-1mu{#1}
	| {#2} \ {#3} | {#4}]}

\def\spaa#1.#2.#3.#4{\langle\mskip-1mu{#1}
	| {#2} \ {#3} | {#4}\rangle}

\title{Off-shell single-top-quark production in the Standard Model Effective Field Theory}

\author[1,2]{Tobias Neumann}

\affil{Department of Physics, Illinois Institute of Technology, Chicago, Illinois 60616, USA}
\affil{Fermilab, PO Box 500, Batavia, Illinois 60510, USA}

\author[1]{Zack Sullivan}

\fancypagestyle{firstpage}{%
	\lhead{}
	\rhead{FERMILAB-PUB-19-119-T, IIT-CAPP-19-01}
}

\begin{document}

\maketitle
\thispagestyle{firstpage}

\begin{abstract}
We present a fully differential and spin-dependent $t$-channel
single-top-quark calculation at next-to-leading order (\NLO{}) in \QCD{} including off-shell effects
by using the complex mass scheme in the Standard Model (\SM{}) and in
the Standard Model Effective Field Theory (\SMEFT{}).  We include all relevant
\SMEFT{} operators at $1/\Lambda^2$ that contribute at \NLO{}
in \QCD{} for a fully consistent comparison to the \SM{} at \NLO{}.
In addition, we include chirality flipping operators that do not
interfere with the \SM{} amplitude and contribute only at
$1/\Lambda^4$ with a massless $b$-quark. 
Such higher order effects are usually captured by considering anomalous
right-handed $Wtb$ and left-handed $Wtb$ tensor couplings. 
Despite their formal suppression in
the \SMEFT{}, they describe an important class of models for new
physics.  Our calculation and analysis framework is publicly available
in \MCFM{}.
\end{abstract}

\clearpage
\tableofcontents

\section{Introduction}

Large statistics data samples from experiments at the \CERN{} Large
Hadron Collider (\LHC{}) provide the opportunity to extract precision
information about the Standard Model (\SM{}), and to look for small
deviations due to new physics that enters at energy scales beyond
direct experimental reach.  $t$-channel single-top-quark production
provides a unique window onto this physics, and has been well-measured
by both the \ATLAS{} 
\cite{Aad:2012ux,Aad:2014fwa,Aaboud:2016ymp,Aaboud:2017pdi} and \CMS{} 
\cite{Chatrchyan:2011vp,Chatrchyan:2012ep,Sirunyan:2016cdg,Sirunyan:2017huu} 
Collaborations at the $\sqrt{S}=7,8$ and \SI{13}{\TeV} runs.  While
the Collaborations refine these measurements to extract limits on
phenomenologically motivated physics beyond the Standard Model (\BSM{})
\cite{Khachatryan:2014iya,Khachatryan:2015dzz,Khachatryan:2016sib,
Aaboud:2017yqf,Aaboud:2017aqp}, the theoretical models they use for
comparison are not generally as precise as the data they are fitting.
This paper describes next-to-leading order (\NLO{}) \QCD{} calculations
of $t$-channel single-top-quark production including off-shell effects
that improve both the \SM{} and Standard Model Effective Field Theory
(\SMEFT{}) predictions of fully differential and spin-dependent
observables.

$t$-channel single-top-quark production probes many aspects of the \SM{}.
Measurement of the cross section provides direct access to the square
of the \CKM{} matrix element
$V_{tb}$ \cite{Stelzer:1997ns,Aaboud:2019pkc}.  The $V-A$ nature of the
production and decay vertices are probed by spin
correlations \cite{Mahlon:1996pn,Mahlon:1999gz}.  Kinematic
distributions, such as the lineshape of the $b$--lepton invariant mass
$m_{bl}$ from the top quark decay products, allow for extraction of
the top quark mass at the \LHC{} \cite{ATLAS:2014baa,Sirunyan:2017huu}.

This process is also a stringent test on the consistency of parton
distribution function (\PDF{}) fits at different orders, and it
directly tests the analytic framework of improved perturbation theory.
Resummation of large logarithms of the top quark mass to the bottom
quark mass leads directly to the introduction a $b$-quark
\PDF{}s \cite{Stelzer:1997ns}.  Through \NLO{} in \QCD{} the
process becomes one of double deep inelastic scattering (\DDIS{})
with two independent scales, where the leading order (\LO{}) process
is $qb \to q^\prime t$ scattering \cite{Stelzer:1997ns}.  Because DIS
data is used to extract the \PDF{}s, when \DDIS{} scale choices are
made, the inclusive $t$-channel cross sections computed at different
perturbative orders should be approximately the same.  A primary
motivation for improving the \SM{} calculation is that this analytic
constraint is strongly violated by recent \PDF{}
sets \cite{Sullivan:2017aiz}.  Once this issue is resolved, this
process could provide insight into \PDF{} transverse momentum
dependence \cite{Abdulov:2018ccp}.

While improving our understanding of \SM{} physics, precision
calculations of single-top-quark production and decay establish a
baseline for controlling the backgrounds to \BSM{} physics.  The final
state of the $t$-channel process is $W$ + $b$ + light jets, where the $W$
can decay to a lepton plus missing energy, and hence is a background
to most new physics models. Deviations in inclusive cross sections or
kinematic distributions are expected in a large class of \BSM{}
physics \cite{Tait:1997fe}.  The spin correlations are especially
sensitive to new physics \cite{Tait:2000sh}, and a number of
observables have been developed in
refs.~\cite{AguilarSaavedra:2010nx,Aguilar-Saavedra:2014eqa,Aguilar-Saavedra:2015yza,Aguilar-Saavedra:2017wpl}
that are separately sensitive to new physics contributions in the
production and decay vertices of the top-quark, respectively.

In the rest of this section we briefly review the state of \SM{}
$t$-channel single-top-quark calculations, explain our focus on the
\SMEFT{} as an extension of the \SM{}, and summarize what we add to the
\SMEFT{} calculations.  In \cref{sec:setup} we describe our setup and the
calculation. We specify the list of \SMEFT{} operators that we use, as
well as our conventions and normalizations. Furthermore, we describe
the steps that we perform to compute and simplify the amplitudes and
provide a list of checks that we have performed. A primary goal of this
study is to allow for a direct improvement of experimental analyses,
and we describe our publicly available implementation in \MCFM{} and
how to use it. In \cref{sec:pheno} we study off-shell and
$W$-boson/neutrino reconstruction effects in the \SM{}. We define
angular observables in the top-quark rest frame that are sensitive
to \SMEFT{} contributions. We then study the impact of off-shell
effects to these distributions, and the effect of higher order
contributions from \QCD{} and stability of the \SMEFT{}.

\paragraph{Towards precise $t$-channel single-top-quark predictions.}

The precision of single-top-quark calculations has largely coincided
with the attempts to discover, and later precisely measure, the
$t$-channel cross section.  Early results focused on the
inclusive \NLO{} cross section with stable on-shell top quarks
\cite{Bordes:1994ki,Stelzer:1997ns,Kant:2014oha}.  Once experimental
backgrounds were better understood, differential calculations were
performed \cite{Harris:2002md,Sullivan:2004ie,Sullivan:2005ar} with a
stable top quark, and the results were used to improve \LO{}
kinematics in showering Monte Carlo programs.

The next step consisted of including the leptonic decay for an
on-shell top quark, preserving full spin correlations, and including
separate \NLO{} corrections in production and
decay \cite{Cao:2004ky,Campbell:2004ch,Schwienhorst:2010je}.  The
on-shell approximation relies on the assumption that off-shell effects
are expected to be small of the order $\Gamma_t / m_t$ inclusively,
where $\Gamma_t$ is the top-quark decay width, and $m_t$ is its
mass. This allows for a significant simplification of the analytical
expressions at the cost of little error for the inclusive observables
used in discovery.

Since the production of the top-quark proceeds through a $b$-quark,
one can distinguish between calculations that either assume an
intrinsic proton $b$-quark content (five-flavor scheme), or not
(four-flavor scheme).  In the latter case, and with a non-zero
$b$-quark mass, predictions were first calculated in the stable
top-quark approximation
at \NLO{} \cite{Campbell:2009ss,Campbell:2009gj}, and then with a decaying
on-shell top quark, retaining spin correlations
\cite{Campbell:2012uf}. This is implemented in \MCFM{} \cite{Campbell:2010ff}.
To this point, all calculations were performed in the on-shell
approximation.

Off-shell effects generally play a role when one considers
differential distributions.  A prime example is the top-quark
invariant mass distribution, where the region above the resonance is
severely underpopulated in the on-shell approximation. It only
receives a tiny contribution when \QCD{} radiation before the
top-quark's decay is clustered with the final state $b$-jet. The
inclusion of off-shell effects was handled for example in an effective field theory
approach \cite{Pittau:1996rp,Beneke:2004km,Falgari:2011qa,Falgari:2010sf},
which is valid only close to the resonance. The first gauge invariant
calculation valid also in the far off-shell region was performed
\cite{Papanastasiou:2013dta,Frederix:2016rdc} in the complex mass scheme 
\cite{Denner:1999gp,Denner:2005fg,Denner:2006ic,Denner:2014zga}, and this
is the approach we follow.

Other calculations include an attempt to improve on fixed order
matching by adding parton shower effects through implementations
in \MCNLO{} for on-shell and off-shell single top
production \cite{Frixione:2005vw,Frederix:2016rdc} and in \PHG{} for
on-shell production \cite{Alioli:2009je}. Analytical resummation has
also been performed on top of the on-shell approximated fixed order
result \cite{Kidonakis:2007ej,Wang:2010ue,Kidonakis:2011wy,Kidonakis:2012rm,Kidonakis:2015wva,Cao:2018ntd,Cao:2019uor}.
Finally, in recent years results at \NNLO{} in \QCD{} have been
published for stable on-shell and for decaying top-quarks
\cite{Brucherseifer:2014ama,Assadsolimani:2014oga,Berger:2016oht,Berger:2017zof}, but these numerical results currently differ by the size of their \NNLO{}
correction terms.

The primary goal of this study is to provide a public implementation
in \MCFM{} of a fully differential spin-dependent prediction for
$t$-channel single-top-quark production and semi-leptonic decay at \NLO{}
including the off-shell effects of initial-final state \QCD{}
interference and of non-resonant interferences.  We demonstrate
in \cref{sec:pheno} that, after cuts, off-shell effects produce
significant shifts in some key experimental observables.
Additionally, we identify kinematic regions and spin observables that
are highly sensitive to the cancellation of soft radiation in
production and decay.  We show that this sensitivity can be hidden by
the on-shell approximation, and thus such regions should be avoided in
analyses relying on fixed order predictions.

\paragraph{New physics in single-top-quark production and decay.} 

Deviations from the \SM{} are frequently modeled using anomalous
couplings (see for example refs.\ 
\cite{Kane:1991bg,AguilarSaavedra:2008zc,Bach:2012fb,Boos:1999dd,Boos:2006af,Boos:2016zmp}) because
they often map directly to experimental observables.  Most recent
studies of \BSM{} physics in the single-top-quark sector
by \ATLAS{} \cite{Aaboud:2017yqf,Aaboud:2017aqp} and \CMS{}
\cite{Khachatryan:2014iya,Khachatryan:2015dzz,Khachatryan:2016sib} use
this approach.  For a recent single-top-quark overview focusing on
measurements of anomalous contributions we refer the reader to
ref.~\cite{Giammanco:2017xyn}.  Without a \UV{} completion, however, it
can be difficult to systematically incorporate and renormalize higher
order perturbative corrections.  It can also be challenging to compare
limits obtained in one experimental data set with limits obtained from
other experiments or data sets since there is neither a systematic
power counting scheme, nor a definite basis of the modifying
structures \cite{Degrande:2012wf} (see also the discussion in
ref.~\cite{Green:2016trm}).

In this paper we take a more systematic approach and parameterize
potentially small deviations in terms of an effective field theory
(\EFT{}) that obeys well-established \SM{} symmetries.  A
classification of all relevant dimension six operators has been
developed in refs.\ \cite{Buchmuller:1985jz,Grzadkowski:2010es}, and
goes under the name Standard Model Effective Field Theory (\SMEFT{}),
see also \cite{Degrande:2012wf}.  Calculations in the \EFT{}
and \SMEFT{} have been performed in abundance at \LO{}, and we can
only cite an excerpt of results, see for example
refs.~\cite{Zhang:2010dr,Zhang:2012cd,Aguilar:2015vsa,AguilarSaavedra:2008zc,
AguilarSaavedra:2008gt,Cao:2006pu,Cao:2007ea,Cao:2015doa,Aguilar-Saavedra:2017nik,AguilarSaavedra:2018nen}. Within
the \EFT{} framework, limits obtained on operators can be compared
directly with limits obtained from $B$-meson decays
\cite{Grzadkowski:2008mf,Fox:2007in}, for example.

To consistently include \NLO{} effects one needs to determine the
renormalization and anomalous dimension matrix of the \SMEFT{}
operators. This framework has been fully developed in a series of
publications \cite{Jenkins:2013zja,Jenkins:2013wua,Alonso:2013hga}. See
also ref.~\cite{Passarino:2016pzb} for a working group report
regarding the importance of \NLO{} corrections of
the \SMEFT{}. Studies including \NLO{} \QCD{} effects for an on-shell
top quark have been performed over the years
\cite{Zhang:2014rja,Franzosi:2015osa,Zhang:2016omx,Drobnak:2010ej}, and recently also for an off-shell top quark 
\cite{deBeurs:2018pvs}, but with a limited set of operators.

The most complete treatment of the \SMEFT{} in single-top-quark
calculations so far has been in ref.~\cite{deBeurs:2018pvs}.  The
authors consider a limited set of three operators contributing as
interference to the \SM{} at \LO{} in \QCD{}, but neglect the
operators that begin contributing at \NLO{} in \QCD{} and operators
that do not interfere with the \SM{}.  Their calculation is also
performed in the complex mass scheme, and includes the effect of
parton showers.  They study different approximations for top-quark
decay in the MadGraph5 framework using MadSpin.  To estimate
uncertainties they consider additional effects at
$\mathcal{O}(1/\Lambda^4)$ that come from squared $1/\Lambda^2$
contributions and from double operator insertions.

We consider six operators that contribute at \NLO{} in \QCD{} at
$1/\Lambda^2$ at the \emph{amplitude} level that are relevant for $t$-channel
production and decay.  In addition, 
we consider a limited set of two color singlet four-fermion operators
with the helicity structure $(\bar{L}L)(\bar{L}L)$ and $(\bar{R}R)(\bar{R}R)$
that are relevant for matching to $W^\prime$ analyses.
We ignore the corresponding color-octet operators, and
$(\bar{L}L)(\bar{R}R)$, $(\bar{L}R)(\bar{L}R)$, and
$(\bar{L}R)(\bar{R}L)$ operators which each have color singlet and
octet contributions. We also ignore four-fermion operators involving the
neutrino and positron. In practice these later operators would only be
relevant for $s$-channel experimental measurements.

Depending on whether the \SMEFT{}
amplitudes interfere with the \SM{}, one obtains effects of order
$1/\Lambda^2$ or $1/\Lambda^4$ at the cross-section level,
respectively.  We do not include effects of $1/\Lambda^4$ at the
amplitude level since these would require a classification of
dimension eight operators for a consistent renormalization
at \NLO{}. Our $1/\Lambda^4$ effects at the cross-section level thus
come from \enquote{squared} amplitudes of order $1/\Lambda^2$ and
constitute only partial effects in view of missing double insertions
and dimension eight operators.

The partial higher order effects we include firstly allow estimating
higher order effects for those operators that already enter at
$1/\Lambda^2$ as an interference with the \SM{}. Secondly, they allow
for a quantification of those operators' effects that only enter
as \enquote{squared} contributions. These are usually neglected
in \EFT{} studies, but would be present in the anomalous couplings
picture. They can be relevant as higher order effects to the
$1/\Lambda^2$ contributions, or under certain model assumptions on the
operator content.

In our study we consider a set of eight operators and all of them are
treated fully consistently at \NLO{} in \QCD{}.  Four of them
contribute at $1/\Lambda^2$ at the cross-section level, and four of
them only enter at $1/\Lambda^4$. Two of the newly considered
operators here only begin to enter at \NLO{} through gluon radiation
and are required for a consistent \NLO{} evaluation since they mix
under renormalization. Note that while the $1/\Lambda^4$ contributions
are partial in the \SMEFT{} expansion, they are still computed
consistently at \NLO{} in \QCD{}. All contributions are implemented
including off-shell top quark effects in the complex mass scheme with
a massless $b$-quark in the five-flavor scheme.  We do not include
all-order effects of parton shower or resummation.

In \cref{sec:setup} we describe the full set of operators, their
relationship to anomalous couplings studies, and technical details of
our calculation.  We show in \cref{sec:pheno} that some commonly
recommended spin-correlation observables
\cite{AguilarSaavedra:2010nx,Aguilar-Saavedra:2014eqa,Aguilar-Saavedra:2015yza,Aguilar-Saavedra:2017wpl}
that are sensitive to these operators are relatively stable to
off-shell effects and QCD radiation, while others are highly sensitive
to soft radiation effects that are not apparent in the on-shell
calculations.


\section{Setup and calculation} \label{sec:setup}

The calculations in this study are performed in the \SM{} and in the \SMEFT{} framework.
The \SMEFT{} is constructed by building higher dimensional operators that respect \SM{} symmetries
out of \SM{} fields. It systematically extends
the \SM{} Lagrangian as a power series in $1/\Lambda$, where
$\Lambda$ is the scale of new physics where the \EFT{} description
breaks down.
\begin{equation}
\mathcal{L}_\SMEFT{} = \sum_i \frac{C_i}{\Lambda^2} \mathcal Q_i + \text{H.c.} \,,
\label{eq:smeft}
\end{equation}
where $\mathcal Q_i$ denote dimension six operators, we add Hermitian
conjugates (H.c.) for non-Hermitian operators with complex Wilson
coefficients $C_i$, and we add Hermitian operators (without H.c.) with
real Wilson coefficients $C_i$.

The \SMEFT{} picture we consider in this paper contrasts with the
phenomenological approach of anomalous couplings, which modify the
$Wtb$ vertex as follows:
\begin{equation}
	-\frac{g_W}{\sqrt{2}}\bar{b} \gamma^\mu ( V_L P_L + V_R P_R) t W^-_\mu - \frac{g_W}{\sqrt{2}} \bar{b} 
	\frac{i\sigma^{\mu\nu}q_\nu}{m_W} (g_L P_L + g_R P_R) t W^-_\mu + \text{H.c.}\,,
	\label{eq:anomc}
\end{equation}
where in the \SM{} $V_L = V_{tb}^*$, $V_R=g_L=g_R=0$, and the
$W$-boson momentum $q$ is chosen to be incoming.  The pictures are connected at
tree level, where the anomalous couplings vertices are generated
by \SMEFT{} operators and can be directly mapped to
them \cite{AguilarSaavedra:2008zc,Bach:2012fb}.  The relations between
the anomalous couplings and our operator Wilson coefficients (defined
further below) are
\begin{align}
 \label{eq:VL} \delta V_L &= \Cone \frac{m_t^2}{\Lambda^2} ,\,\text{where } V_L = 1 + \delta V_L\,,\\
 \label{eq:VR} V_R &= \Ctwo{}^* \frac{m_t^2}{\Lambda^2}\,, \\
 \label{eq:gL} g_L &= -4\frac{m_W m_t}{\Lambda^2} \cdot \Cfour\,, \\
 \label{eq:gR} g_R &= -4 \frac{m_W m_t}{\Lambda^2} \cdot \Cthree{}^*\,,
\end{align}
where $m_W$ is the $W$-boson mass, and $m_W = \frac{1}{2} g_W v$ has
been used to derive this equivalence.  Note that the minus sign for
$g_L$ and $g_R$ is different from the literature. It depends on the choice of
treating the momentum $q_\nu$ in \cref{eq:anomc} as incoming or outgoing and
the sign convention in the covariant derivative. We treat all momenta as
incoming in this study and adopt the sign convention from the \SMEFT{}
literature (see further below).

When considering the leading $1/\Lambda^2$ contributions from
dimension six operators to single top observables, only $V_L$ and
$g_R$ contribute as interference to the \SM{} amplitude. The
contributions generated by $V_R$ and $g_L$ flip a $b$-quark chirality
and do not interfere with the \SM{} amplitude for a massless
$b$-quark. As such, in the \SMEFT{} they only enter at order
$1/\Lambda^4$ as \enquote{squared} contributions from non-\SM{}
helicity amplitudes. Operator double insertions would contribute to
the amplitudes at the same order, where renormalization requires the
inclusion of dimension eight operators and a determination of their
anomalous dimension matrix. Dimension eight operators are beyond the
scope of this study, and so we do not include them or any double
insertions.

While formally suppressed in \SMEFT{}, the couplings $V_R$ and $g_L$ are
relevant for the study of new charged vector currents ($W^\prime$
bosons) or scalars ($H^\pm$
bosons) \cite{Sullivan:2002jt,Duffty:2012rf,Drueke:2014pla} and are
strongly constrained experimentally despite their suppression.  This
is possible due to the strong spin correlations in single-top-quark
production, and a large set of observables highly sensitive to \SM{}
deviations.

To maintain a direct coupling to experiment, we follow a
hybrid approach. In the first part we include all dimension
six \SMEFT{} operators that are relevant at \NLO{} in \QCD{}, and
enter at order $1/\Lambda^2$ at the cross-section level. This allows
for a fully consistent evaluation of \SMEFT{} effects and comparison
of Wilson coefficients extracted from different experiments when
higher order effects can be neglected.

In an enhanced mode we include all contributions of order
$1/\Lambda^2$ in the \emph{amplitudes} from dimension six operators of
the \SMEFT{}. The additional operator contributions do not interfere
with the \SM{} and contribute only as the operator insertions squared,
or in interference with other \SMEFT{} contributions. This leads to
$1/\Lambda^4$ effects at the cross-section level.  This enhancement
serves two purposes.  It allows for a \NLO{} \QCD{} mapping to
anomalous coupling studies under the assumption that dimension eight
operators can be ignored.  And it allows for a systematic
determination of whether a given observable is sensitive to
higher-order corrections in the EFT.  If one wishes to obtain
consistent limits on Wilson coefficients and compare them with other
sources, one should not be sensitive to $1/\Lambda^2$ and
$1/\Lambda^4$ contributions at the same time. For that purpose one can
run the analysis using only the $1/\Lambda^2$ contributions and then
compare with results obtained when including the partial $1/\Lambda^4$
contributions.

\paragraph{\SMEFT{} operators.}

As shown in \cref{eq:smeft}, all operators $\mathcal{Q}$ come with a
Wilson coefficient $C$ and a power of $1/\Lambda^2$.  The operators
that contribute at $1/\Lambda^2$ in \NLO{} \QCD{} as interference to
the \SM{} amplitude are
\begin{align} 
	\mathcal{Q}_{\varphi q}^{(3,33)} &=  \frac{1}{2} y_t^2 (\varphi^\dagger i \overset{\leftrightarrow}{D_\mu^I} 
	\varphi) 
	(\bar{Q}_L 
	\gamma^\mu \tau^I Q_L)\,, \\
	\mathcal{Q}_{uW}^{33} &= y_t g_W (\bar{Q}_L \sigma^{\mu\nu} \tau^I t) \tilde{\varphi} W^I_{\mu\nu}
	\,, \\
	\mathcal{Q}_{uG}^{33} &= y_t g_s (\bar{Q}_L \sigma^{\mu\nu} T^A t) \tilde{\varphi} G^A_{\mu\nu} \,, \\
	\Qeight &= \mathcal{Q}_{qq}^{(3,1133)} = ( \bar{q}_L \gamma_\mu \tau^I q_L)  (\bar{Q}_L \gamma^\mu \tau^I Q_L) \,, 
\end{align}
where $Q_L$ is the third generation left handed \SU(2) doublet
$(t_L,b_L)$ and $q_L$ the first generation doublet $(u_L,d_L)$.  Here
$y_t=m_t\sqrt{2}/v$ is the real-valued top-quark Yukawa coupling,
$g_W$ is the electroweak coupling, and $g_s$ is the strong
coupling. The operators $\Qone$ and $\Qeight$ are Hermitian, have real
Wilson coefficients, and no Hermitian conjugate is added to the sum
in \cref{eq:smeft}. We also add the second generation operator
$\mathcal{Q}_{qq}^{(3,2233)}$ with the same real Wilson
coefficient.The operator $\Qsix$ modifies the $\bar{t}tg$ Feynman
rule vertex and enters only at \NLO{} in \QCD{}.

Our notation, sign conventions, and operator basis follows that of the \SMEFT{}
literature in
refs.~\cite{Grzadkowski:2010es,Jenkins:2013zja,Jenkins:2013wua,Alonso:2013hga}.
This sign convention is \emph{different} from the one used in other $t$-channel single-top-quark
\SMEFT{} studies \cite{Zhang:2010dr,Zhang:2012cd,Zhang:2014rja,Zhang:2016omx,deBeurs:2018pvs}.  Both conventions exist in FeynRules model files \cite{Alloul:2013bka}.
We use a plus sign for minimal coupling in the covariant derivative
$D_\mu = \partial_\mu + i g X_\mu$ for gauge fields $X$ and a
corresponding minus sign in the gauge field strength tensor
$X_{\mu\nu} = \partial_\mu X_\mu + \partial_\nu X_\mu - g X_\mu X_\nu$
consistent with SMEFTsim package \cite{Brivio:2017btx}, whereas the
dim6top package \cite{AguilarSaavedra:2018nen} uses the opposite
convention.

At $1/\Lambda^4$ there are additional dimension six operators
\begin{align}
	\mathcal{Q}_{\varphi u d }^{33} &= y_t^2 (\tilde{\varphi}^\dagger i D_\mu \varphi) ( \bar{t} \gamma^\mu b)\,, \\
	\mathcal{Q}_{dW}^{33} &= y_t g_W (\bar{Q}_L \sigma^{\mu\nu} \tau^I b ) \Phi W^I_{\mu\nu}\,, \\
	\mathcal{Q}_{dG}^{33} &= y_T g_s(\bar{Q}_L \sigma^{\mu\nu} T^A b) \Phi G^A_{\mu}\nu\,, \\
	\Qnine &= \mathcal{Q}_{ud}^{(1,1331)} + \mathcal{Q}_{ud}^{(1,3113)} = (\bar{d} \gamma_\mu u) ( \bar{t} \gamma^\mu 
	b) + (\bar{u} \gamma_\mu d) ( \bar{b} \gamma^\mu t) \,,
\end{align}
where the third operator $\Qseven$ only contributes at \NLO{}
in \QCD{} and modifies the $\bar{b}bg$ vertex. For the Hermitian
operator $\Qnine$ no Hermitian conjugate is added to the sum in
\cref{eq:smeft}, and we also add the corresponding second generation operator 
with the same Wilson coefficient.

\paragraph{Operator mixing and running.}

The operator pair $\Qthree, \Qsix$ has nonzero anomalous dimension and
mixes according to
\begin{equation}
	\mu_X \frac{\mathrm{d}}{\mathrm{d}\mu_X} \begin{pmatrix}
	 \Qsix \\ \Qthree
	\end{pmatrix} = \frac{\alpha_s}{4\pi} C_F \begin{pmatrix}
	1 & 0 \\
	2 & 2 
	\end{pmatrix}
	\begin{pmatrix}
	\Qsix \\ \Qthree
	\end{pmatrix}\,,
	\label{eq:mixing}
\end{equation}
where $\mu_X$ is a renormalization scale that is independent of
the \QCD{} renormalization scale. The pair $\Qfour, \Qseven$ mixes
analogously under renormalization \cite{Alonso:2013hga}.  

We renormalize the Wilson coefficients in the $\overline{\text{MS}}$
scheme following
$C_i^\text{bare} = Z_{ij} C_j(\mu)$, where
$$Z_{ij} = 1
+ \frac{\alpha_s}{4\pi} \frac{\gamma_{ij}}{2\epsilon}\,.$$ 
A factor of $(4\pi)^\epsilon / \Gamma(1-\epsilon)$ is absorbed into
the definition of $\alpha_s$ and from here on we define
$a_s \equiv \alpha_s/(4\pi)$. We set the renormalization point of the
Wilson coefficients to the same value as the \QCD{} renormalization
point, as both contributions are probed at the same scale. The effect
of the running of $C_i$ has been studied in
refs.~\cite{Zhang:2016omx,Zhang:2014rja} and can be used to evolve the
Wilson coefficients to the scale $\Lambda$ or to some lower scale for
comparisons.

We renormalize the top-quark mass and wavefunction in the complex mass
scheme with complex mass on-shell conditions.  The \SM{}
renormalization constants receive additional contributions from
$\Qsix$. 
We confirm the \SMEFT{} mass and wavefunction
renormalization constants in ref.~\cite{Zhang:2014rja} by computing
the top-quark 1PI self energy with complex mass on-shell
renormalization conditions and find
\begin{align}
	\mu_0 &= (1 + a_s \delta_m) \mu_t\,, \\
	\delta_m &= \left(\frac{\mu^2}{\mu_t^2}\right)^\epsilon C_F \left( -\frac{3}{\epsilon} -4 + 
	\Real\, \Csix \frac{m_t \mu_t}{\Lambda^2} C_F \left( \frac{12}{\epsilon} + 4 \right)
	 \right)\,, \\
	Z_\Psi &= (1 + a_s \delta Z_\Psi)\,, \\
	\delta Z_\Psi &= \left(\frac{\mu^2}{\mu_t^2}\right)^\epsilon C_F \left( -\frac{3}{\epsilon} - 4 +
	\Real\, \Csix  \frac{m_t \mu_t}{\Lambda^2} C_F \left( \frac{6}{\epsilon} + 2 \right) +
	\Imag\, \Csix  \frac{m_t \mu_t}{\Lambda^2} C_F\, i\gamma_5 \left( \frac{6}{\epsilon} + 2	\right)
	  \right)\,.
\end{align}
Here $\mu_t^2 = m_t^2 - i \Gamma_t m_t$ is the squared complex
top-quark mass. One power of $m_t$ is part of the operator
normalization and is kept real. Note that in ref.~\cite{Zhang:2014rja}
the covariant derivative is defined with a different sign convention which
flips the sign of the Wilson coefficient $\Csix$ relative to our results.

Special care has to be taken for $\Imag \Csix$, which receives a
wavefunction renormalization contribution proportional to
$\gamma_5$. We obtain it by adding an additional counterterm to the
top-quark 1PI self energy proportional to $\gamma_5$ and demanding
that the propagator keeps its tree-level form throughout higher
orders. For the $\gamma_5$ contribution this is analogous to the \SM{}
on-shell condition of having residue $i$ for the renormalized
propagator.

Note: Because we are examining an off-shell top quark, the
non-resonant diagrams include contributions from $\bar{b} b A$ and
$\bar{b} b Z$ vertices, where $A$ is the photon field. The
corresponding \SM{} Feynman rules receive contributions from $\Qfour$
in the \SMEFT{} and renormalize the gluon contributions from $\Qseven$
at \NLO{}. In this case also the operator $\mathcal{Q}_{dB}^{33} = y_t
g_B ( \bar{Q} \sigma^{\mu\nu} b) \Phi B_{\mu\nu}$ must be included to
renormalize $\Qseven$. The anomalous dimension matrix follows the
operator renormalization group running
\begin{equation}
\mu_X \frac{\mathrm{d}}{\mathrm{d}\mu_X} \begin{pmatrix}
\mathcal{O}_{dG}^{33} \\ \mathcal{O}_{dW}^{33} \\ \mathcal{O}_{dB}^{33}
\end{pmatrix} = \frac{\alpha_s}{4\pi} C_F \begin{pmatrix}
1 & 0 & 0 \\
2 & 2 & 0 \\
-2/3 & 0 & 2 
\end{pmatrix}
\begin{pmatrix}
\mathcal{O}_{dG}^{33} \\ \mathcal{O}_{dW}^{33} \\ \mathcal{O}_{dB}^{33}
\end{pmatrix}\,.
\label{eq:mixingtwo}
\end{equation}
We include this operator $\Qten$ only for the renormalization of
$\Qseven$ and set its Wilson coefficient to zero afterwards, since it
only contributes to the non-resonant amplitudes.

\subsection{Technical implementation and checks}
\label{sec:technical}

We consider the process $u(p_1) + b(p_2) \to \nu(p_3) + e^+(p_4) +
b(p_5) + d(p_6)$ at \NLO{} in \QCD{} in the complex mass scheme
including off-shell interference effects and non-resonant
contributions required by gauge invariance. The complex mass scheme
introduces a squared complex top-quark mass $\mu_t^2 = m_t^2 - i
m_t\Gamma_t$ to the otherwise real valued top-quark mass in the
Lagrangian.\footnote{Note that the \EFT{} operators above have been
normalized with the real-valued on-shell top quark Yukawa coupling
$y_t=\sqrt{2} m_t /v$.}  We also work in the five-flavor scheme and
set $m_b=0$. We compute all results at the amplitude level in the
spinor helicity formalism in the 't~Hooft-Veltman scheme and avoid any
ambiguities related to $\gamma_5$ in dimensional regularization and
thus treat it in the naive dimensional regularization approach.

\begin{figure}\centering
	\subfloat[resonant]{ \includegraphics[height=4cm]{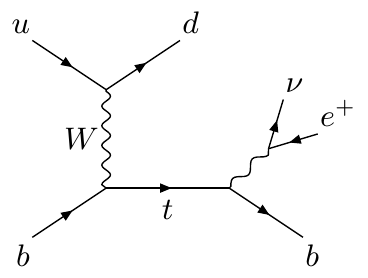}}
	\subfloat[non-resonant]{ \includegraphics[height=4cm]{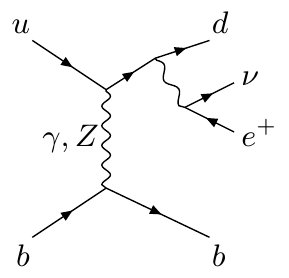}}
		\subfloat[non-resonant]{ \includegraphics[height=4cm]{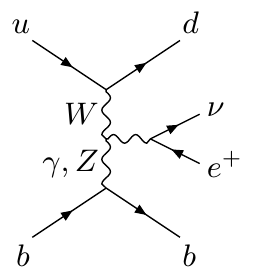}}
	\caption{Sample Feynman diagrams of resonant and non-resonant contributions at \LO{}.}
	\label{fig:diagsLO}
\end{figure}

\begin{figure}\centering
	\subfloat[non-resonant \SMEFT{}]{ \includegraphics[height=4cm]{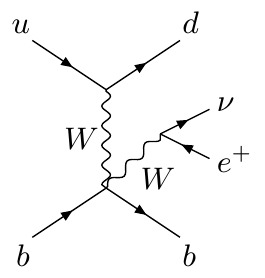}}
	\subfloat[off-shell resonant]{ \includegraphics[height=4cm]{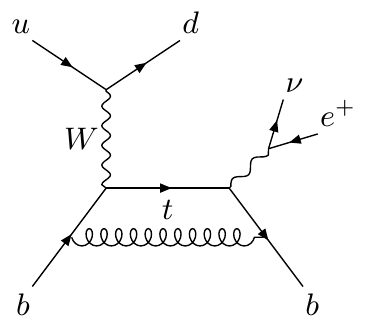}}
	\caption{(a) Example non-resonant contribution in the \SMEFT{} at \LO. (b) One-loop resonant diagram 
	with production-decay interference.  }
	\label{fig:diagstwo}
\end{figure}

\begin{figure}\centering
	\subfloat[production]{ \includegraphics[height=4cm]{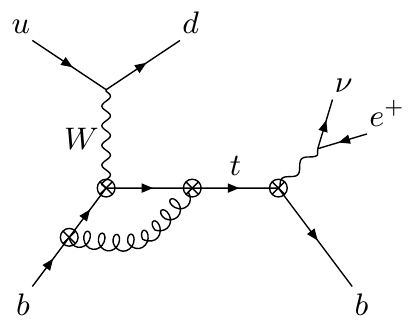}}
	\subfloat[decay]{ \includegraphics[height=4cm]{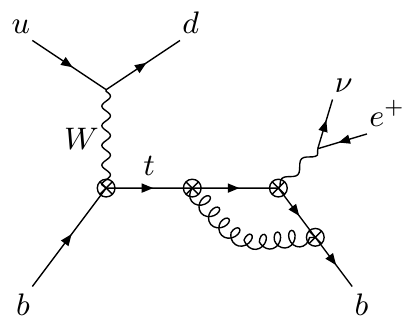}}
        \subfloat[self-energy]{ \includegraphics[height=4cm]{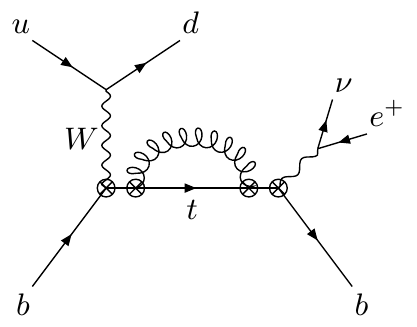}}
	\caption{NLO virtual contribution to the (a) production vertex, (b)
decay vertex, or (c) top-quark self-energy.  Each crossed circle represents a possible
\SMEFT{} operator insertion.}
	\label{fig:diagsvirt}
\end{figure}

To obtain a gauge invariant result with an off-shell top quark,
requires the inclusion of both resonant and non-resonant
contributions. We include all such contributions and show a partial
sample of the diagrams
in \cref{fig:diagsLO,fig:diagstwo,fig:diagsvirt}. In addition to \QCD{}
corrections, we allow for exactly one \SMEFT{} operator insertion in
each diagram (at positions denoted by the crossed circles in \cref{fig:diagsvirt}).
We do not include the $W$+2 jets contributions that have a gluon
exchange at tree level. These diagrams are separately gauge invariant,
do not interfere with our contributions through \NLO{}, and are
considered a background that can be computed fully independently.

Apart from gluon radiation, in the \SM{} and throughout \NLO{} only
two helicity amplitudes contribute. The first amplitude only
encompasses left-handed particles and a right-handed positron. Its
predominant contribution is from the resonant diagrams, which are
purely left-handed charged current mediated. It also receives
non-resonant contributions from $Z$-boson and photon exchanges.  The
second amplitude with flipped $b$-quark helicities comes purely from
non-resonant pieces. When \SMEFT{} operators are added, one has
additional helicity amplitudes where either one of the $b$-quark
helicities is flipped ($\Qtwo,\Qfour,\Qseven$), or the helicities of
the light quark line are flipped together with the initial-state
$b$-quark helicity ($\Qnine$).

The one-loop amplitudes we want to compute and simplify have large
tensor ranks that are further increased with the \SMEFT{}
contributions.  Their evaluation with a standard framework like
FeynCalc \cite{Mertig:1990an,Shtabovenko:2016sxi} and
Passarino-Veltman reduction, for example, would be prohibitively
difficult due to the size of resulting and intermediate
expressions. We instead develop our own setup in
Mathematica \cite{Mathematica} which performs the tensor reduction
with dimensional shift relations \cite{Tarasov:1996br}. We implement
the dimensionally shifted integrals with increased propagator powers
in terms of standard one-loop master integrals by means of integration
by parts reduction performed with Kira \cite{Maierhoefer:2017hyi}. The
scalar one-loop integrals are evaluated with \QCDLoop{}
2.0 \cite{Ellis:2007qk,Carrazza:2016gav}.  The few necessary Feynman
diagrams are generated with \QGRAF{} \cite{Nogueira:1991ex} and
translated into initial \FORM{} \cite{Vermaseren:2000nd} code
with \DIANA{} \cite{Tentyukov:1999is} to output Mathematica
code. Feynman rules are generated using \LANHEP{}
\cite{Semenov:1996es,Semenov:2002jw,Semenov:2008jy,Semenov:2014rea}
and checked by-hand, as well as compared with
refs.~\cite{Dedes:2017zog,Romao:2012pq}. We make use of the
Mathematica packages \SaM{} \cite{Maitre:2007jq} and FeynCalc
\cite{Mertig:1990an,Shtabovenko:2016sxi} for debugging purposes. The 
simplification of large expressions is accelerated enormously with the
multivariate polynomial greatest common divisor implementation in
Fermat \cite{Fermat}, and it is used through an interface to
Mathematica \cite{mmaFermat}.

A major part of our calculation involves the reduction of spinor
(helicity) chains to a set of basis structures, as outlined in
ref.~\cite{Denner:2005fg}, where the elements of the minimal set are
referred to as \enquote{standard matrix elements.} We extended these
reduction prescriptions to spinor chains of type
left-right \enquote{$\langle \cdots \rangle$} and
right-left \enquote{$[ \cdots ]$}, which appear in our \SMEFT{}
contributions.  In principle, a reduction to one spinor master
structure is possible for the \SM{} helicity
amplitudes \cite{Denner:2005fg}; and a reduction to two master
structures can be performed for the amplitudes with one flipped
$b$-quark helicity (left-right and right-left types).  In practice, we
balance the number of structures we use against the number of terms
produced when we express the coefficients in terms of kinematic
invariants.

We begin by reducing all spinor combinations to a set of $59$
structures. To achieve a full reduction to two master structures one
can directly write down a set of 58 linearly independent equations in terms
of nine Lorentz-invariants \cite{Denner:2005fg}. The use of Lorentz-invariants to
parameterize the kinematics enforces one additional Gram determinant
constraint \cite{Byckling:1971vca}, which leaves 57 independent
equations.  This system of equations is highly complicated, but can be solved with the aid
of Kira \cite{Maierhofer:2018gpa,Maierhoefer:2017hyi} for example. The
resulting expressions are huge and do not directly to lead to
simplifications when inserted in the amplitudes.  Instead we follow
the suggestion of ref.~\cite{Denner:2005fg}, and perform the reduction
using only equations that do not introduce additional denominator
structures.  For example, we express the \SM{} amplitude with
left-handed $b$-quarks and contributions from
$\Qone, \Qthree, \Qsix, \Qeight$ to the same helicity configuration in
terms of five spinor chains.  We reduce to a larger basis set for the
amplitudes with flipped quark helicities, since the equations would
either introduce additional complicated denominator structures, or do
not lead to simpler final results.
 
Both resonant and non-resonant contributions contain box diagrams that
naively lead to huge expressions. While expressing the results in
terms of scalar box integrals in six dimensions removes some of the
cancellations between the box and triangle diagrams, we do not use a
basis that lends itself to simple expressions for the loop amplitudes.
In order to deal with leftover spurious cancellations that eventually will
impair numerical stability we implement a simple rescaling scheme
stability control mechanism.

This stability control works as follows: For each phase space point we
evaluate our matrix elements twice --- in double precision, and again
in double precision where all dimensionful quantities are rescaled by
a constant. Taking into account the dimensionality of the matrix
element, we then divide out the constant and check how many digits
agree to get an estimate for the numerical precision of the result. We
find that for precision runs the integration eventually focuses its
sampling on numerically unstable points and the integration becomes
unstable. If the stability check fails with less than four digits
precision left, we reevaluate them using the \QD{}
library \cite{libqd}, which implements twice the precision of \IEEE{}
doubles (approximately 32 decimal digits) using two double precision
variables. This is faster than an evaluation with full \IEEE{} quad
precision and allows for Fortran compilers without such quad precision
support.

We directly compute all amplitudes for single-top-quark production. The
matrix elements for single-top-antiquark production are obtained by
crossing the single-top-quark matrix elements after applying a CP
transformation. After the CP transformation one crosses the light
quarks and reidentifies electron and neutrino particle labels to
obtain the single-top-antiquark result.  The CP transformation itself
introduces sign flips for the imaginary parts of the Wilson
coefficients, which we take into account, but leaves the other parts
unaffected since we assume $V_{tb}$ is real.

In order to maintain a connection to \PDF{} fits, we implement the use
of double deep inelastic scattering (\DDIS{})
scales \cite{Stelzer:1997ns} in \MCFM{}.  We label the DIS momentum
transfer between the light-quark line and the $b$-quark line as
$Q^2=-q^2$. We then set the renormalization and factorization scales
for the light-quark line to $\mu^2=Q^2$, and for the $b$-quark line
$\mu^2 = Q^2 + m_t^2$. The implementation of \DDIS{} scales at \NLO{}
is a non-trivial effort, since light line and heavy line corrections
have to be handled separately, and Catani-Seymour dipole contributions
have also to be accounted for with the right scales.

As part of our calculation we compute the decay width $t\to W b$
at \NLO{} including the \SMEFT{} operators, with an on-shell $W$-boson
and a massless $b$-quark. This is consistent with the complex mass
scheme at \NLO{}. We follow the steps of ref.~\cite{Campbell:2004ch}
to perform the necessary real emission phase space integrals. In
addition to the integrals listed in ref.~\cite{Campbell:2004ch} table
I, we find that three additional finite phase space integrals $\langle
y \rangle$, $\langle z \rangle$ and $\langle y z \rangle$ are
necessary. We compute these using the Mathematica package
HypExp \cite{Huber:2005yg}.

\paragraph{Crosschecks.}
We compute Feynman rules in the \SM{} and the \SMEFT{} with
both \LANHEP{} and by hand, ensuring proper relative signs with the
help of ref.~\cite{Romao:2012pq}. Our \SMEFT{} Feynman rules agree
with those in ref.~\cite{Dedes:2017zog}. The
relative signs between $\Qthree$ and $\Qsix$, and $\Qfour$ and
$\Qseven$ are fixed by the anomalous dimension matrix in
ref.~\cite{Alonso:2013hga}, and we agree with this through our
operator renormalization.

We compared our results analytically to the \SMEFT{} tree level matrix
elements printed in ref.~\cite{Zhang:2010dr}, eq. (27). We also compared
our results with the \SMEFT{} \NLO{} decay width results in eq. (120) of
ref.~\cite{Zhang:2014rja} as well as with the mass and wavefunction renormalization
constants. We fully agree with all those results when taking into account
the different sign convention used in the covariant derivative for
the \SMEFT{} operator definition.

Using our anomalous couplings parametrization in \cref{eq:anomc} with the
$W$-boson momentum defined to be incoming, we find
a \emph{global} sign difference for $\Qthree$ and $\Qfour$ when comparing our results against the Protos
code \cite{AguilarSaavedra:2008gt,AguilarSaavedra:2008zc}. This difference
is explained by the different momentum convention in Protos, where the momentum is outgoing.
We do agree on the relative signs between real and imaginary parts, so this difference
is indeed purely a global sign of $\Qthree$ and $\Qfour$.

Another check on our signs that eliminates a consistency problem with defining
momenta as incoming or outgoing in the Feynman rules is as follows.
The needed Feynman rules in off-shell single top production at \NLO{}
from $\Qthree$ and $\Qsix, \Qseven$ all have a linear momentum
dependence, and one might argue that a sign difference can appear
be because we define particle momenta as incoming. However, because
we are considering off-shell effects, we must introduce the operator
$\Qfour$, which adds an additional contribution from the momentum
independent $W^+ W^- \bar{b} b$ vertex. The relative sign between this
contribution and the contributions from $\Qfour$ and $\Qseven$
at \NLO{} is set by the cancellation between the poles for \UV{}
renormalization and \IR{} subtraction. As we are overall consistent
with the signs in ref.~\cite{Dedes:2017zog}, this fixes the signs of
both $\Qfour$ and $\Qseven$. Since $\Qthree$ and $\Qsix$ have the same
structure and mixing, it also fixes their signs. This is the
first calculation to include the operators $\Qfour$ and $\Qseven$
at \NLO{} (in the off-shell process), so this check is new.

We explicitly checked \QCD{} gauge invariance for our amplitudes ---
analytically for pole terms and numerically for finite pieces. We note
that all amplitudes contributing to $\Qsix$ and $\Qseven$ are
separately gauge invariant. As part of our setup we reproduce
the \SM{} \NLO{} decay width, which is calculated in detail in
refs.~\cite{Campbell:2004ch,Jezabek:1988iv}. We extensively compare
our off-shell \SM{} calculation to on-shell results for
compatibility. We check the proper cancellation of poles
between real and virtual corrections by checking independence of the
$\alpha$ parameter in the \MCFM{} Catani-Seymour dipole
implementation \cite{Catani:1996vz,Nagy:1998bb,Nagy:2003tz} to the
per-mille level. We also check that all infrared singular limits of
the real emission amplitude are approached correctly as predicted by
the Catani-Seymour dipole terms.

\subsection{Implementation in MCFM-8.3}

Our results are implemented in the release version 8.3
of \MCFM{}. Here we describe the user-visible modifications
of \MCFM{}.  The code allows one to directly and easily reproduce the
plots in the following phenomenology section.  We implement $b$
tagging, top-quark and $W$ reconstruction, as well as preconfigured
histograms for the most common observables in the \SMEFT{} and spin
correlation studies. Our implementation provides an easy analysis
framework to perform further studies.

Dynamical double deep inelastic scattering scales can be
consistently used at \NLO{} by setting \texttt{dynamicscale} to `DDIS'
and \texttt{scale}$=$\texttt{facscale} to 1d0. In this way the
momentum transfer along the $W$-boson $Q^2$ is used as the scale for
the light-quark-line corrections $\mu^2=Q^2$, and $\mu^2=Q^2+m_t^2$ for
the heavy-quark-line corrections. These scales are also consistently
used for the non-resonant contributions, with \QCD{} corrections on the
$ud$-quark line, and separate \QCD{} corrections on the bottom-quark
line.

The new block \enquote{Single top SMEFT, nproc=164,169} in the input
file governs the inclusion of \SMEFT{} operators and corresponding
orders.  The scale of new physics $\Lambda$ can be separately set, and
has a default value of $\SI{1000}{\GeV}$.  The flag \texttt{enable
1/lambda4} enables the $1/\Lambda^4$ contributions, where operators
$\Qtwo, \Qfour, \Qseven$ and $\Qnine$ can contribute for the first
time.  For the non-Hermitian operators we allow complex Wilson
coefficients.  We also have a flag to disable the pure \SM{}
contribution, leaving only contributions from \SMEFT{} operators
either interfered with the \SM{} amplitudes or as squared
contributions at $1/\Lambda^4$.  This can be used to directly and
quickly extract kinematical distributions and the magnitudes of
pure \SMEFT{} contributions.

To allow for easier comparisons with previous anomalous couplings
results, and possibly estimate further higher order effects, we allow
for an anomalous couplings mode at \LO{} by enabling the corresponding
flag.  The relations between our operators and the anomalous couplings
are the same as in \cref{eq:VL,eq:VR,eq:gL,eq:gR}.

The analysis/plotting routine is contained in the file
`\texttt{src/User/nplotter\_ktopanom.f}', where all observables
presented in this study are implemented, and the $W$-boson/neutrino
reconstruction is implemented and can be switched on or off. With this
one can directly reproduce all the phenomenological results in this
study.


\section{Phenomenology} \label{sec:pheno}

In this section we examine kinematic distributions in off-shell single-top-quark
production and decay in both the \SM{} and the \SMEFT{}.
We begin by examining the effects of a
$W$-boson / neutrino reconstruction on the top-quark reconstruction. We
then study a set of angular observables in the top-quark rest frame
for the \SM{} before we focus our attention on \SMEFT{}
contributions. We address the importance of unique \NLO{} perturbative
corrections to the \SMEFT{} contributions compared to using \LO{}
predictions with \SM{} $K$-factors. We also show the behavior of the
operators $\Qsix$ and $\Qseven$ that only enter at \NLO{} and are
shown here for the first time for the full process.

Our set of cuts is given in \cref{tab:loosecuts}. We require at least
one $b$-jet and one non-$b$-jet, but also allow for a third jet of
either kind. We refer to the leading non-$b$-jet as the spectator
jet. On top of these cuts, experimental anomalous couplings studies in
$t$-channel production select exactly two jets and have further cuts
on the rapidities of the $b$ and spectator jet
\cite{Aaboud:2017yqf,Aaboud:2017aqp,Khachatryan:2014iya,Khachatryan:2015dzz,Khachatryan:2016sib}. We find that these 
additional cuts decrease the acceptance, but do not alter any of our conclusions here.

\begin{table}
	\centering
	\caption{Applied cuts at a center of momentum energy $\sqrt{s}=\SI{13}{\TeV}$, $m_t^\text{O.S.}=\SI{173}{\GeV}$, 
		$\mu_X=\mu_R=\mu_F$ set to \DDIS{} scales. $W$-boson and top-quark reconstruction are as described in the text.}
	\vspace*{1em}
	\begin{tabular}{l|c}
		Jets & $p_\text{T,jet} > \SI{30}{\GeV}$, $|\eta_\text{jet}| < 4.5$, $R_\text{jet} = 0.4$ \\
		& 	at least one $b$-jet and one non-$b$-jet (spectator)  \\
		Lepton & $p_\text{T}^l > \SI{25}{\GeV}$, $\abs{\eta^l} < 2.5$ \\
		Neutrino & $p_\text{T}^\nu > \SI{30}{\GeV}$ \\
	\end{tabular}
	\label{tab:loosecuts}
\end{table}

Our default choice of renormalization and factorization scales for the
off-shell results are the \DDIS{} scales, where for the light-quark
line the momentum transfer $Q^2$ to the $b$-quark line is used,
$\mu^2=Q^2$, and for the $b$-quark line $\mu^2 = Q^2 + m_t^2$. It has been shown that the \DDIS{} scales
lead to small perturbative corrections in inclusive
observables \cite{Sullivan:2004ie,Stelzer:1997ns}.  We confirm that
the difference between using a fixed scale $\mu^2=m_t^2$ and
the \DDIS{} scales is tiny for most \NLO{} accurate observables, even
differentially. For \LO{} observables like the subleading $b$ or
subleading light quark jet transverse momentum, which only enter
through the real emission, the scale choices lead to significant
differences.

We note that in on-shell results, we use the fixed scale $m_t$ which
is used throughout the literature.  While a comparison between
off-shell results with \DDIS{} scales and on-shell results with $m_t$
as a scale is not on precisely equal footing, we consider the \DDIS{}
scales to be an improvement over the current calculations that do not
allow for this natural scale choice.  While the \DDIS{} paradigm
formally breaks down at \NNLO{} in \QCD{}, interference between the
light- and heavy-quark lines is expected to be
small \cite{Brucherseifer:2014ama,Berger:2016oht}.  In addition, there
remains an analytic relationship to DIS in PDF
fits \cite{Sullivan:2017aiz} that is directly constrained by the
consistency of the calculation of \DDIS{}.

We use \CTFOURTEEN{} parton distribution functions
(\PDF{}s) \cite{Dulat:2015mca} at the corresponding perturbative
orders with a value of $\alpha_s^\NLO{}(m_Z)=0.118$ at \NLO{}, and at
$\alpha_s^\LO{}(m_Z)=0.13$ at \LO{}. The \CKM{} matrix is chosen to be
diagonal and all other parameters have recent PDG values as
implemented in \MCFM{}-8.3. The top-quark width is evaluated at the
corresponding perturbative order for $t\to W b$ at the fixed scale
$m_t$ and takes into account the \SMEFT{} contributions at \LO{}
and \NLO{}.

\subsection{Off-shell and $W$-reconstruction effects in the Standard Model.}
\label{subsec:recon}

It is well known that in fixed order perturbation theory colored
resonances are sensitive to soft radiation \cite{Aeppli:1993rs}.  At
higher orders in perturbation theory soft and collinear parton
configurations between virtual corrections and real emission
corrections cancel in the singular limit.  However, configurations
approaching the soft/collinear limits are still present. In our case,
the top-quark is reconstructed from a reconstructed $b$-jet and
$W$-boson.  Depending on whether such radiation configurations get
clustered with the $b$-jet, and whether the radiation is
produced before the resonance or in its decay, one can observe a mass
enhancement or diminution. 

Assuming the top-quark is in the on-shell approximation, the
cancellation between virtual corrections and real emission is pinched
to the phase space with an on-shell reconstructed top-quark.  Having
an off-shell top quark makes the approach of the cancellation
explicit, with large positive and negative contributions around
$\simeq m_t \pm \Gamma_t$.  To obtain a smooth invariant mass
distribution near the peak one can either choose a larger binning with
radius $\simeq \Gamma_t$ or include all-order effects through parton
shower or resummation.

A further complication is that experimental analyses have to use a
reconstruction scheme for the leptonically decaying $W$-boson.  The
neutrino's transverse component can be derived by requiring the
event's transverse momentum to be vanishing.  On the other hand, the
longitudinal component is completely unknown and needs to be
reconstructed.  This reconstruction induces a smearing, not just of
the $W$-boson, but also of the reconstructed top quark.  We follow the
most recent \ATLAS{} study on anomalous coupling
contributions \cite{Aaboud:2017yqf} and reconstruct the neutrino's
four-vector by requiring that the invariant mass of the
neutrino-electron system in the top-quark decay equals the on-shell
$W$-boson mass.  With this condition one has either two real solutions
or two complex solutions for the neutrino's longitudinal component.
In the former case the solution closer to zero is taken.  For the
latter case of complex solutions, the neutrino's transverse component
is rescaled by $0.9$ until a real and positive solution is found.

\begin{figure}
	\includegraphics{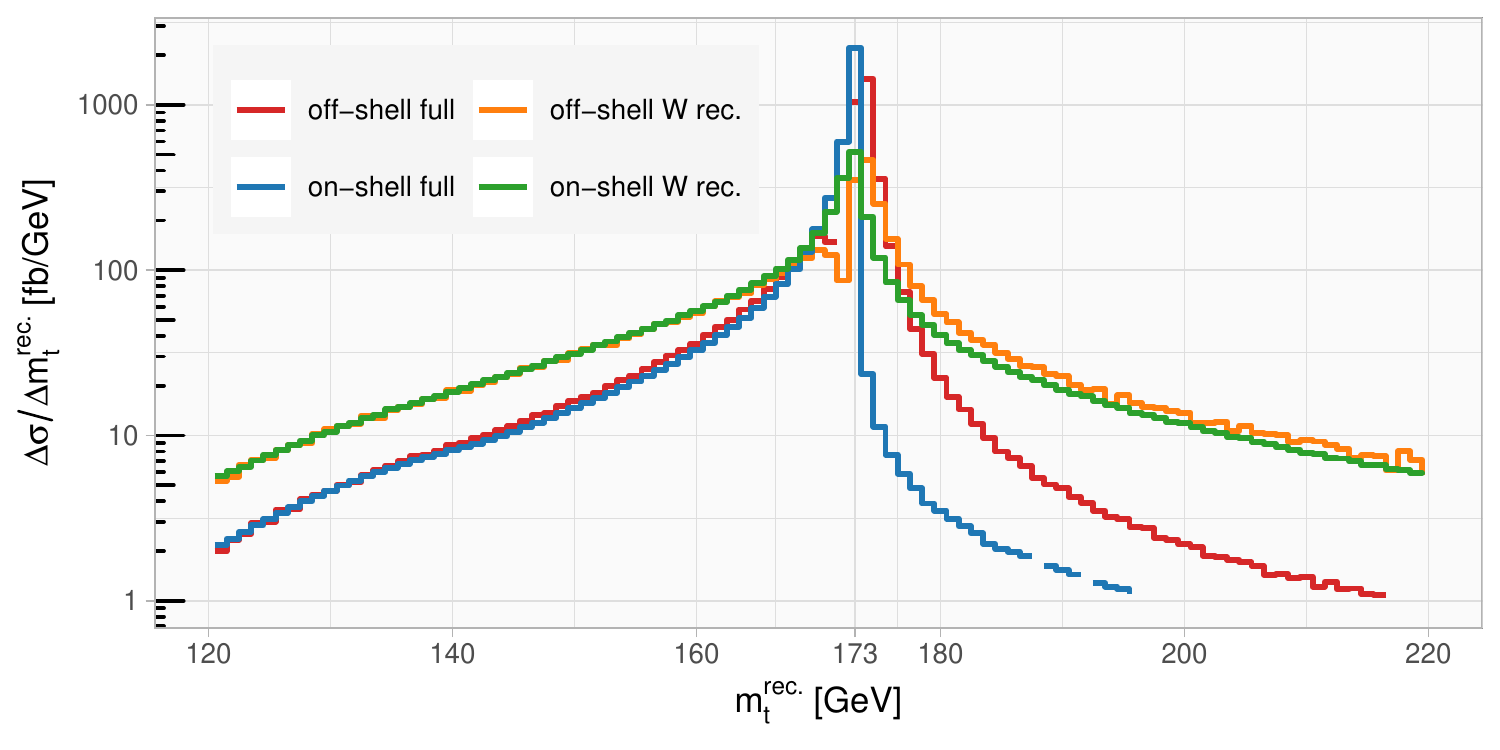}
	\caption{Reconstructed top-quark mass distribution for an on-shell and off-shell top quark at \NLO{}. The full 
		result denotes reconstruction through the full $W$-boson and $b$-jet four vectors. The \enquote{W rec.} lines
		denote the reconstruction of the neutrino's longitudinal component.}
	\label{fig:smmt}
\end{figure}

As a result of the neutrino reconstruction the reconstructed top-quark
invariant mass distribution gets smeared, and the aforementioned problem
is somewhat ameliorated, although not fully removed. We show this
in \cref{fig:smmt}, where we compare the reconstructed top-quark
invariant mass distribution using the full neutrino four momentum to
using the reconstructed neutrino (smeared). For comparison the
on-shell distributions are also shown. The full off-shell result
receives large negative (not shown on the logarithmic scale) and
positive contributions close to the resonance. These are smeared by
using the reconstructed neutrino, but one can still see a noticeable
dip just below $m_t$. These off-shell effects in comparison to the
on-shell approximation are well known
\cite{Papanastasiou:2013dta,Falgari:2011qa,Falgari:2010sf}.

We note that the previous off-shell \SM{} calculation in the complex
mass scheme in ref.~\cite{Papanastasiou:2013dta} with an on-shell
$W$-boson seems to obtain a smooth $m_t$ distribution
with \SI{1}{\GeV} bins without applying any smearing procedures.  This
behavior could be due to the way the $W$-boson is handled, leading to
a smoothing effect, although no indication of how to treat the $W$-boson
decay is given in their study.  It is also conceivable that since the
affected region is not well-defined in fixed order perturbation
theory, their use of a different subtraction procedure leads to a
differently distributed result there.

\begin{figure}
	\centering\includegraphics{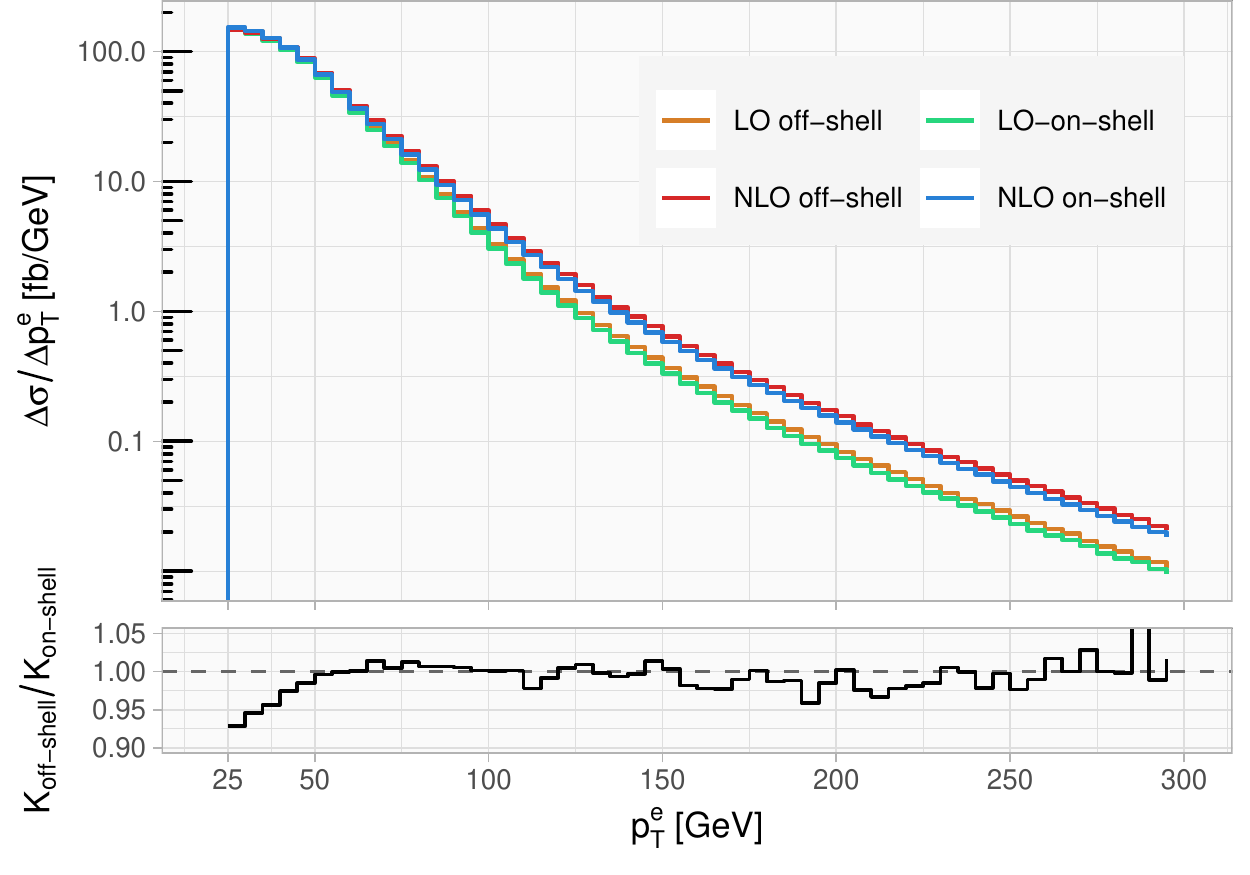}
	\caption{Positron transverse momentum distributions for the top-quark on-shell approximation and for the 
		off-shell top quark at \LO{} and \NLO{}. The lower panel shows the ratio of the \NLO{}/\LO{} $K$-factors from 
		the off-shell and on-shell results.
	}
	\label{fig:smpte}
\end{figure}

\begin{figure}
	\centering\centering\includegraphics{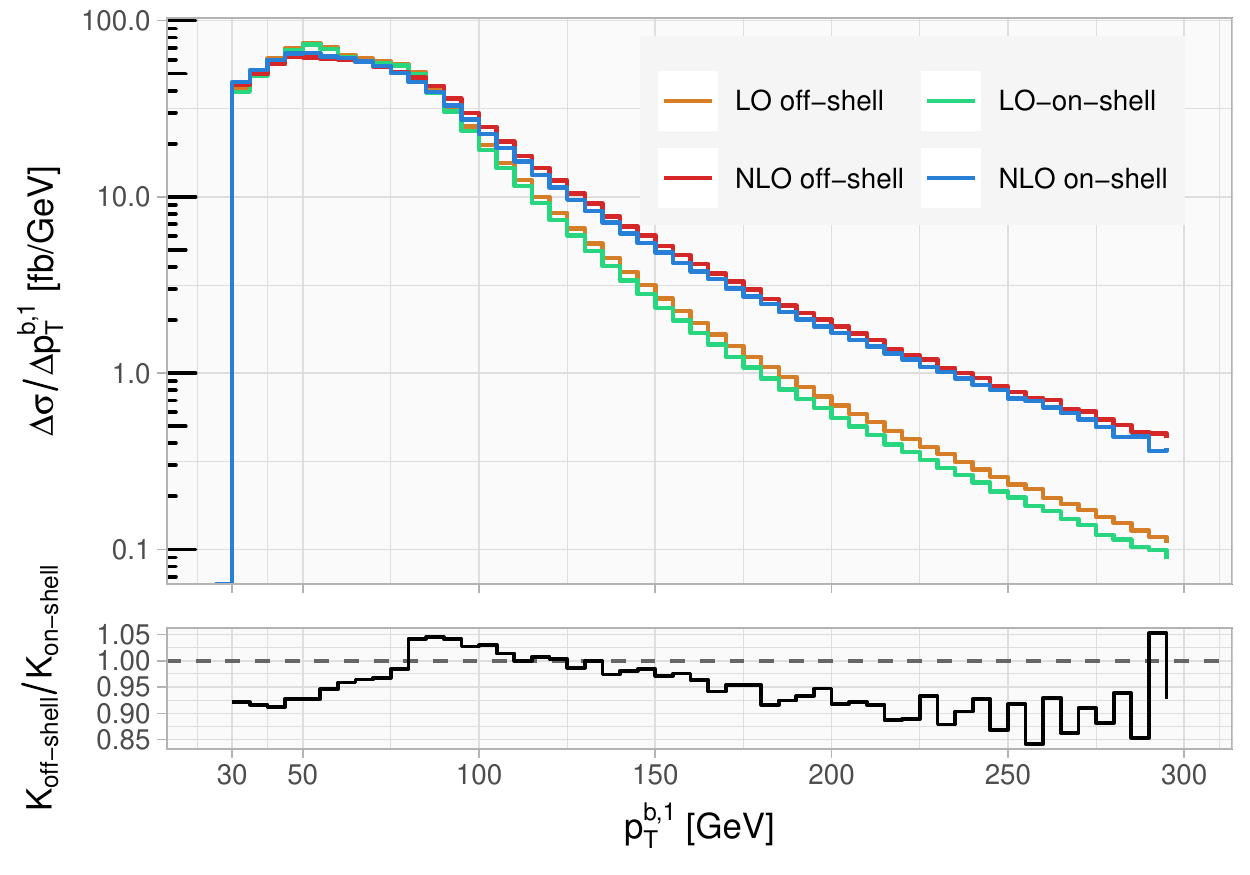}
	\caption{Leading $b$-jet transverse momentum distributions for the top-quark on-shell approximation and for the 
		off-shell top quark at \LO{} and \NLO{}. The lower panel shows the ratio of the \NLO{}/\LO{} $K$-factors from 
		the 
		off-shell and on-shell results.	}
	\label{fig:smptb1}
\end{figure}

The off-shell effects we consider are important for experimental
observables.  In \cref{fig:smpte} we show the positron transverse
momentum distribution, which is sensitive to soft \QCD{} radiation
corrections only through the recoil of the $W$ boson.  The off-shell
distribution is up to $\sim 15\%$ harder at \SI{300}{\GeV} compared to
the on-shell result, while corrections at low $p_T$ are at the few
percent level.  The $K$-factor
($\sigma_{\mathrm{NLO}}/\sigma_{\mathrm{LO}}$) in an on-shell
calculation and the $K$-factor in the off-shell calculation differ by
at most a few percent.  Similar
corrections can be seen in the leading $b$-jet transverse momentum
distribution in \cref{fig:smptb1}. There, off-shell effects at \NLO{}
are about $5-10\%$ in the tail. The ratio of the $K$-factors
for the off-shell and on-shell production is not flat, and shows
deviations with up to $\sim 10\%$.

\begin{figure}
	\centering\includegraphics{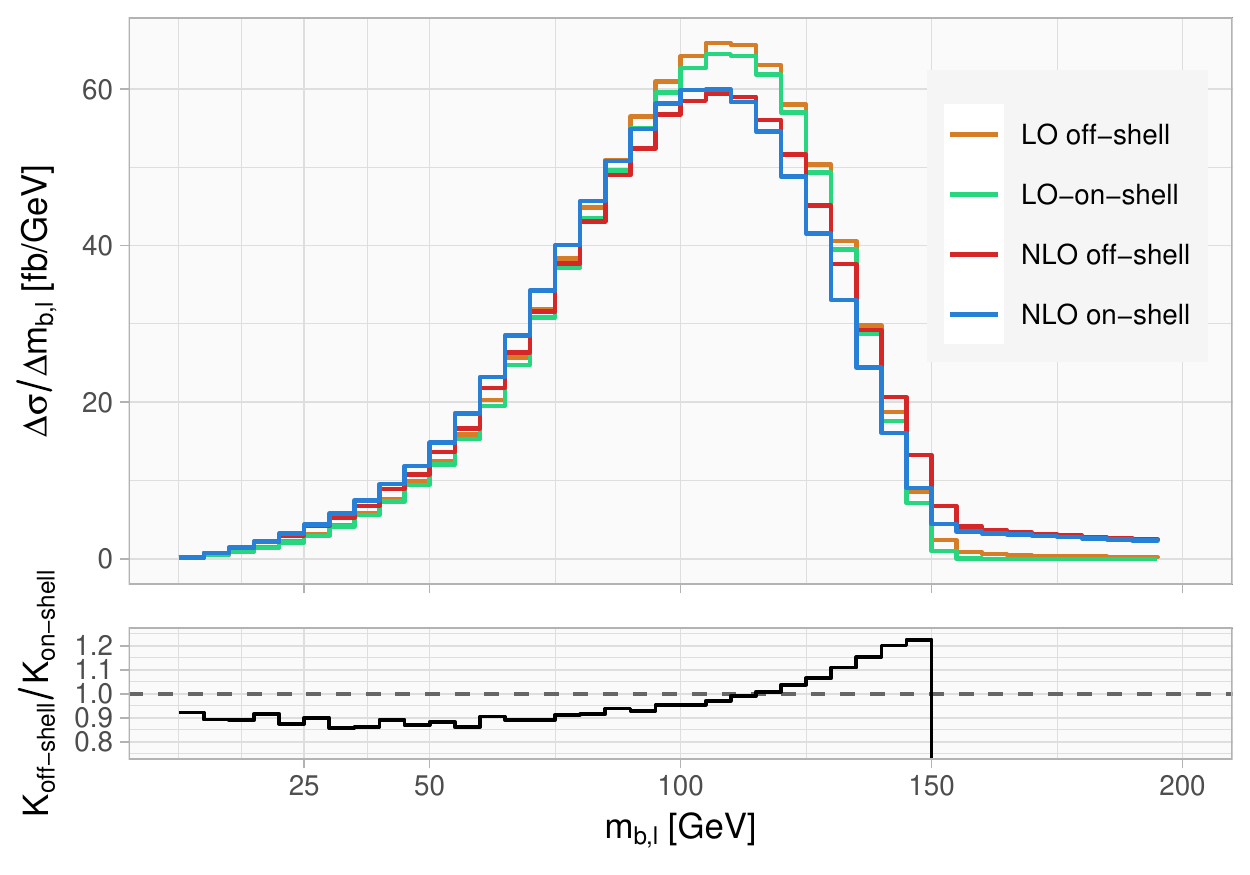}
\caption{Invariant mass of the positron plus leading $b$-jet system for the top-quark on-shell approximation and for 
the 
	off-shell top quark at \LO{} and \NLO{}. The lower panel shows the ratio of the ratio of the \NLO{}/\LO{} 
	$K$-factors from the off-shell and on-shell results.
}
\label{fig:smmbl}
\end{figure}

Deviations in the distributions of these top-quark decay products will
have a significant effect on \LHC{} measurements of the top-quark mass.
To avoid neutrino reconstruction uncertainties, it is common
practice to fit the top-quark mass based on the line shape of the
$b$-jet/lepton invariant mass
$m_{bl}$ \cite{ATLAS:2014baa,Sirunyan:2017huu}.  In \cref{fig:smmbl}
we observe that off-shell effects lead to a large $\sim 20\%$ shift in
the $m_{bl}$ line shape close to the kinematic endpoint.  In a template
fit this effect is similar to a few GeV shift in the top-quark mass,
though the difference might be partially ameliorated by further final
state showering.

\subsection{Angular observables in the top-quark rest frame}

Apart from common kinematical distributions like transverse momenta,
rapidities and invariant masses, $t$-channel single-top-quark analyses
are characterized by their use of angular distributions.  In
particular, the angle between the leading non-$b$ jet and the lepton
from the top-quark decay is strongly correlated in the top-quark rest
frame \cite{Mahlon:1996pn,Mahlon:1999gz}, and this is used in part to
identify the $t$-channel state \cite{Sullivan:2005ar}.  Any
modification (other than scaling) of the production or decay vertices
is expected to be observed as a modification of one or more angular
observables \cite{Aguilar-Saavedra:2014eqa,AguilarSaavedra:2010nx}.
While most of the angles are well-behaved under the inclusion of
off-shell effects from fixed order perturbation theory, we find that
one angle is highly unstable to soft radiation.


The first set of angles we examine is sensitive to operators that
modify the production of the top quark.  A coordinate system is
established by using the direction of the spectator quark $\vec{p}_j$
in the top-quark rest frame to define a corresponding axis $\hat{z}$.
A second axis $\hat{y}$ is defined by the direction orthogonal to the
plane made by the spectator quark and the initial-state light-quark,
while the third axis $\hat{x}$ is defined by requiring the coordinate
system to be right-handed \cite{Aguilar-Saavedra:2014eqa}. The
direction of the initial-state quark is taken to be that of the proton
beam that shares the same sign of rapidity as that of the spectator
jet.
\begin{equation}
\hat{z} = \frac{\vec{p}_j}{|\vec{p}_j|},\quad \hat{y} = \frac{\vec{p}_j \times \vec{p}_q}{|\vec{p}_j \times 
\vec{p}_q|},\quad \hat{x} = \hat{y} \times \hat{z} \label{eq:prodangles}\,.
\end{equation}
We refer to angles of the lepton in the top-quark rest frame with
respect to these axes as $\cosxprod, \cosyprod, \coszprod$.

The second coordinate system we consider is sensitive to the decay
vertex, and starts with the direction $\vec{q}$ of the $W$-boson in
the top-quark rest frame as one axis $\hat{q}$. The second axis
$\hat{N}$ is orthogonal to the plane defined by $\hat{q}$ and the
top-quark spin direction, as implemented by the spectator quark
direction $\vec{s}_t$ in the top-quark rest frame. The last axis
$\hat{T}$ is again defined by the right-handedness of the coordinate
system \cite{AguilarSaavedra:2010nx}:
\begin{equation} 
\hat{q} = \frac{\vec{q}}{|\vec{q}|},\quad \hat{N} = \frac{ \vec{s}_t \times \vec{q} }{|\vec{s}_t \times
\vec{q}|},\quad \hat{T} = \hat{q} \times \hat{N} \label{eq:decayangles}\,.
\end{equation}
From this basis we construct the angles between the lepton in the
$W$-boson rest frame with respect to these three axes.  We refer to
them as $\coslstar$ for the angle with respect to the $\hat{q}$ axis,
and $\coslN$, $\coslT$ with respect to $\hat{N}$ and $\hat{T}$,
respectively.  In addition to the angles described here, one can find
angles between the $\hat{N}$ and $\hat{T}$ axes, and projections of
the lepton in the $W$ boson rest frame onto the $\hat{N}$-$\hat{T}$
plane being used in analyses.  We have examined those projection
angles, but do not find interesting results regarding the \SMEFT{}
contributions.

\paragraph{Discussion of angular observables.}

First we note that neutrino reconstruction has a noticeable impact on
most of the observables in the top-quark rest frame.  This is expected
as the top-quark rest frame has a direct dependence on the neutrino
four-momentum.  Since the reconstruction procedure we use is based on
an experimental algorithm described in \cref{subsec:recon}, we do not
comment further on angular differences due to other reconstruction
procedures.  Both on-shell and off-shell results that follow use a
reconstructed neutrino.

It was previously observed \cite{Sullivan:2005ar} that, after
cuts, going from \LO{} to \NLO{} had little effect on the \SM{}
angular distributions used to measure $t$-channel single-top-quark
production.  When comparing off-shell distributions to on-shell
results, this similarity in most angular distributions is maintained.
For example, for $\cosxprod$ the $K$-factor ratio is $\simeq 0.97 -
0.98$ and flat within $1-2\%$ of integration noise. In the \SM{},
off-shell effects have rather uniform impact.  With one notable
exception, the largest deviations in shape we find are $\sim 10\%$ in
$\cosyprod$ and $\coslstar$ which are modestly relevant when
considering backgrounds to \SMEFT{} operators. For completeness we
include corresponding plots with $K$-factor ratios
in \cref{fig:cosxprodsm,fig:cosyprodsm,fig:coszprodsm,fig:coslTsm,fig:coslstarsm}
in \cref{sec:appendix}.

\begin{figure}
	\centering\includegraphics{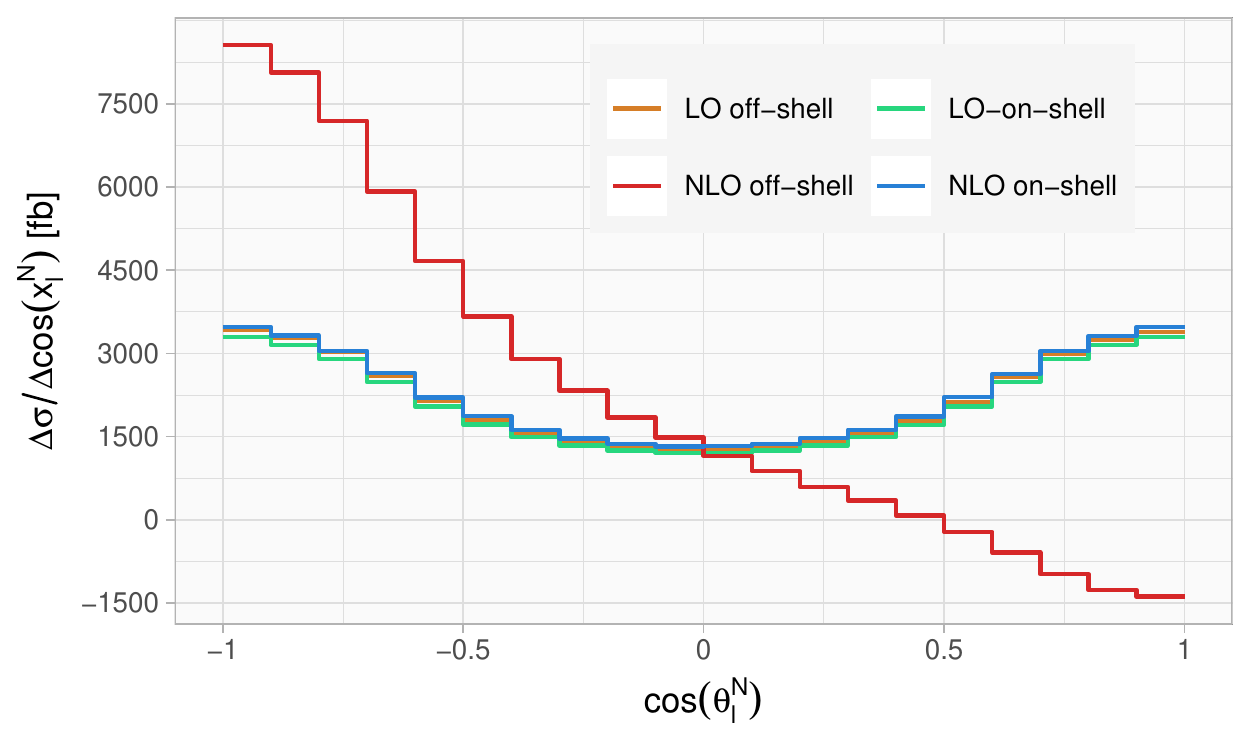}
	\caption{Angular distribution $\coslN$ at \LO{} and \NLO{} for an on-shell and off-shell top quark. The \NLO{} 
	off-shell result becomes unphysical and indicates a strong sensitivity to the cancellation of soft radiation.}
	\label{fig:coslN}
\end{figure}

One angular distribution that is currently used in experimental
analyses \cite{Aaboud:2017aqp} demands further discussion.  The
angular distribution of $\coslN$ becomes unphysical at \NLO{} for an
off-shell top quark in fixed order perturbation theory,
see \cref{fig:coslN}.  This is because the top-quark invariant mass
distribution is not well-defined close to the resonance, where soft
radiation from production and decay cancels. It turns out
that the angle $\coslN$ is highly sensitive to this cancellation and
the cross section prediction becomes negative for $\coslN \gtrsim 0$,
which is compensated by an according increase for $\coslN \lesssim 0$.

When using this observable at \LO{} or with the on-shell approximation
one seemingly does not have this issue, since no negative cross
section is observed. But the intrinsic sensitivity to the cancellation
of soft radiation in this angle is merely hidden.  Effects from
resummation and parton showers can change the distribution drastically.
As such we do not recommend to use this angle for precision studies in
the \SM{} and the \SMEFT{}. One must be especially careful in the latter
case, as \SMEFT{} operators can modify gluon radiation. A
resummation or parton shower without taking into account the \SMEFT{}
operators might lead to incorrect conclusions or limits.

\subsection{\SMEFT{} contributions}

We now turn our attention to modifications of the angular
distributions by \SMEFT{} operators in the off-shell \NLO{} \QCD{}
calculation.  We distinguish how \SMEFT{} operators modify the
distributions compared to \SM{} \NLO{} effects.  We also discuss the
importance of higher order corrections in the \EFT{} for the
consistent extraction of limits on \SMEFT{} Wilson coefficients.
Operators that begin to enter at $1/\Lambda^2$ receive corrections at
order $1/\Lambda^4$.  To obtain universal results one needs to make
sure that these higher order \EFT{} corrections are small.

We limit ourselves to some representative examples here, as a detailed
study of all operators and their effects on all observables used in
single-top studies is beyond the scope of this paper. Additional
observables can easily and quickly be predicted with our published
code. We only show off-shell results here and present $K$-factors for
them, if applicable. We do not display $\Qone$ in our plots since this
operator is just a rescaling of the \SM{} results with an effective
modification of $V_{tb}$. We consider the case where just one Wilson
coefficient is non-zero and choose Wilson coefficients of one or $i$, with a
scale $\Lambda = \SI{1000}{\GeV}$. This choice is not very important
here, except for the consideration of higher order effects
$1/\Lambda^4$. Otherwise, our presentation of the pure (\SM{}
subtracted) and normalized \SMEFT{} contributions divides out the
Wilson coefficient.

\paragraph{\NLO{} and $1/\Lambda^4$ effects.} 

We start with a discussion of higher order effects in \QCD{} and in
the \SMEFT{}. It has already been pointed out in
ref.~\cite{deBeurs:2018pvs} that \emph{inclusively} the $K$-factors
for the \SMEFT{} contributions are different from the \SM{} $K$-factor
broadly by $10-25\%$ depending on the operator combination.  We show
below that \emph{differentially} this worsens somewhat.  For the
actual distributions used to constrain \BSM{} physics, \NLO{} \QCD{}
corrections to the \SMEFT{} operators are essential.

\begin{figure}
	\centering\includegraphics{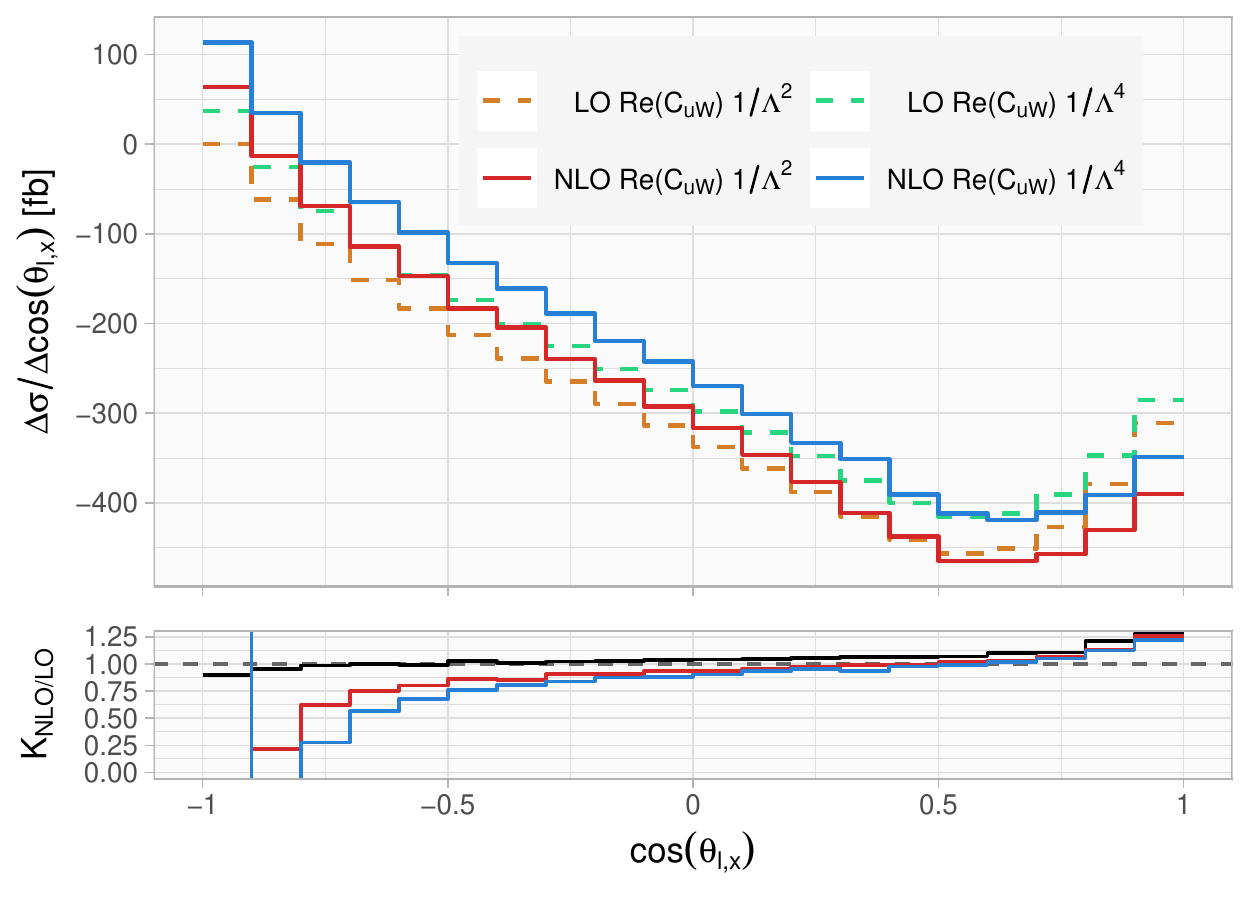}
	\caption{Distribution of $\cosxprod$ for the pure \SMEFT{} contribution with $\Real\,\Cthree = 1$, 
	$\Lambda=\SI{1000}{\GeV}$. Shown are results at \LO{} and \NLO{} in \QCD{} and at $1/\Lambda^2$ in the \SMEFT{} as 
	well as with higher order effects $1/\Lambda^4$.}
	\label{fig:cosxprodreo3}
\end{figure}

We begin with one of the most important operators $\Qthree$, that
leads to the richest phenomenological structure.
In \cref{fig:cosxprodreo3} we show the pure \SMEFT{} contribution
(with the \SM{} contribution subtracted) at \LO{} and \NLO{} in \QCD{}
to get an impression of the perturbative corrections. We also include
effects of order $1/\Lambda^2$ and additionally of order $1/\Lambda^4$
to show the impact of higher order \EFT{} corrections. In the top
panel the absolute corrections are shown and in the bottom panel we
show $K\equiv \NLO{}/\LO{}$ as a representation of the perturbative
corrections.  For comparison we have also included the $K$-factor for
the \SM{} contribution itself in black.

\begin{figure}
	\centering\includegraphics{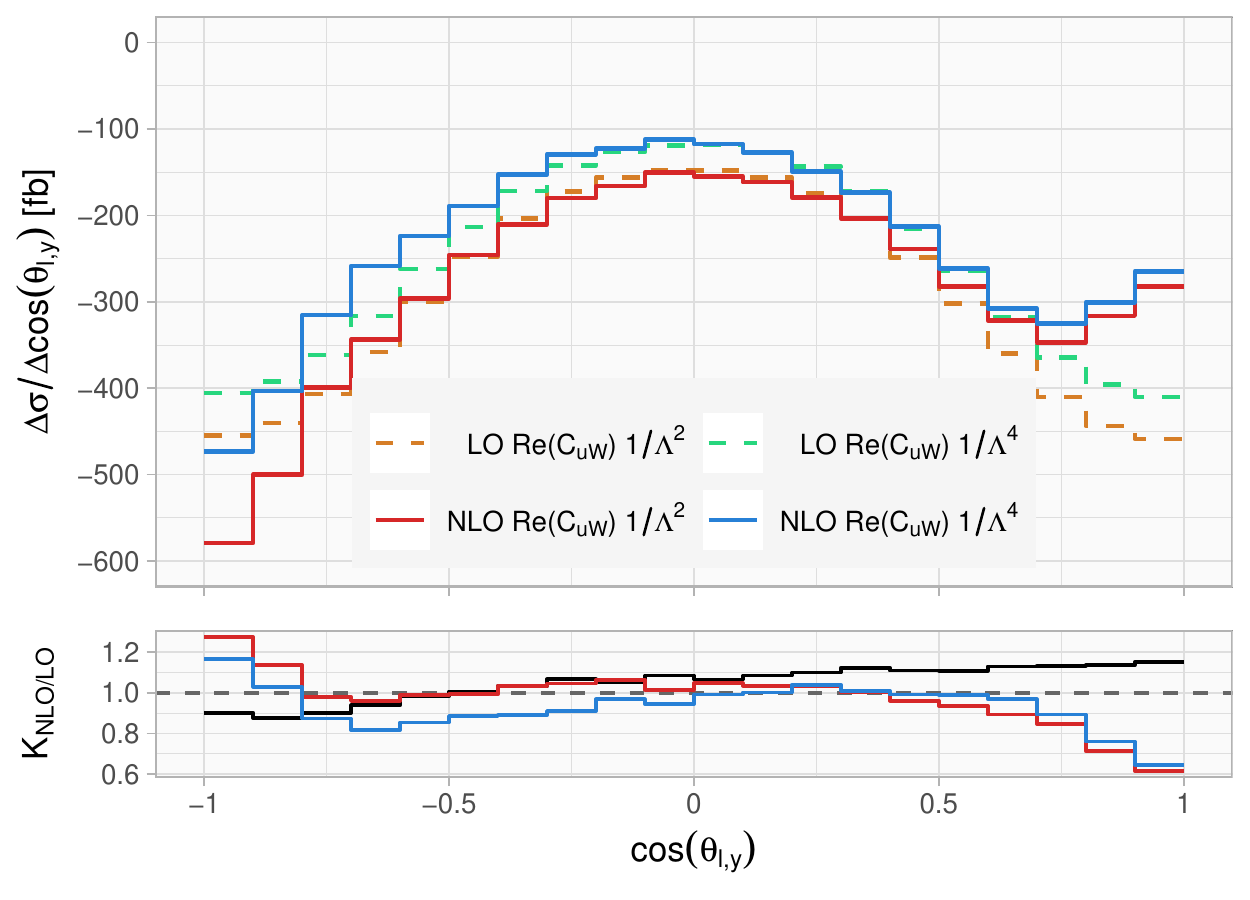}
	\caption{Distribution of $\cosyprod$ for the pure \SMEFT{} contribution with $\Real\,\Cthree = 1$, 
		$\Lambda=\SI{1000}{\GeV}$. Shown are results at \LO{} and \NLO{} in \QCD{} and at $1/\Lambda^2$ in the \SMEFT{} 
		as 
		well as with higher order effects $1/\Lambda^4$.}
	\label{fig:cosyprodreo3}
\end{figure}

The $K$-factor for the \SMEFT{} contribution is not flat, and
unique \NLO{} \QCD{} corrections are indeed important, especially for
the region of $\cosxprod \simeq - 1$. The impact of the $1/\Lambda^4$
corrections on the $K$-factor is moderate in size, apart from the
first bins. This effect reduces for a smaller Wilson coefficient or a
larger scale $\Lambda$, but might have to be considered depending on
the analysis.  Differentially the corrections are an important effect
to consider, but the size of the contributions at $\cosxprod \simeq
-1$ are small in comparison to the other regions. We show the same
operator contributions for $\cosyprod$ in \cref{fig:cosyprodreo3}.
Generally \NLO{} corrections are sizable differentially and important
to correctly capture the shapes.

\paragraph{The imaginary part of $\Qthree$.}

While for most of the $1/\Lambda^2$ contributions with Wilson
coefficients of one, the higher order \EFT{} effects seem to be
moderate, this is not the case for the imaginary part of $\Qthree$'s
Wilson coefficient, $\Imag\,\Cthree$.  It it claimed in the
literature that the imaginary part of $\Qthree$ does not contribute at
$1/\Lambda^2$ \cite{deBeurs:2018pvs}.  We do not
find this to be true (with our set of cuts). For the stable-top approximation
one indeed does not have enough linearly independent four-vectors contracted with the Levi-Civita tensor,
so the contribution vanishes
\cite{Cirigliano:2016nyn}, but this no longer holds for a decaying top-quark.
The Wilson coefficient $\Imag\,\Cthree$ enters the composite observable $\delta_{-}$ 
\cite{Boudreau:2013yna,Cirigliano:2016nyn,Alioli:2017ces} as measured for example
by \ATLAS{} \cite{Aad:2015yem}.

\begin{figure}
	\centering\includegraphics{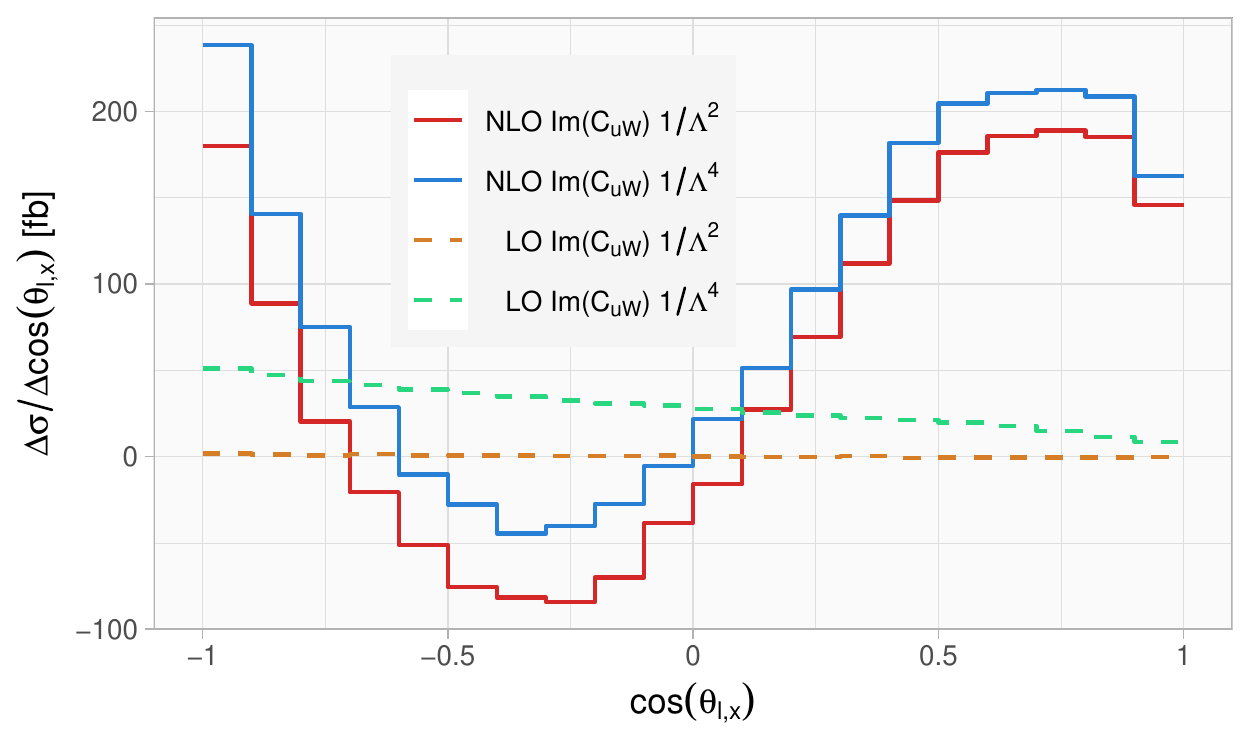}
	\caption{Distribution of $\cosxprod$ for the pure \SMEFT{} contribution with $\Imag\,\Cthree = 1$, 
	$\Lambda=\SI{1000}{\GeV}$. Shown are results at \LO{} and \NLO{} in \QCD{} and at $1/\Lambda^2$ in the \SMEFT{} as 
	well as higher order effects $1/\Lambda^4$. }
	\label{fig:cosxprodimo3}
\end{figure}

\begin{figure}
	\centering\includegraphics{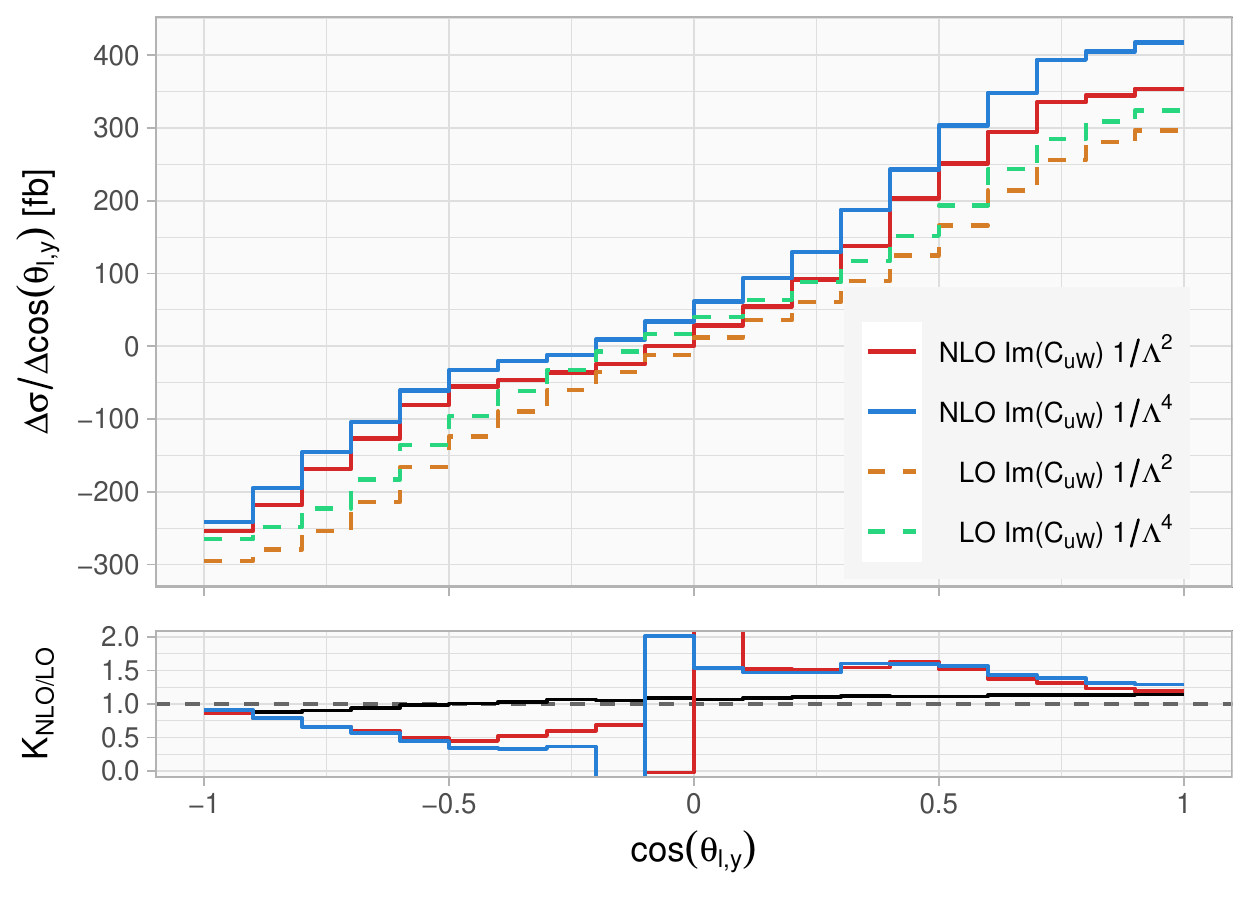}
	\caption{Distribution of $\cosyprod$ for the pure \SMEFT{} contribution with $\Imag\,\Cthree = 1$, 
		$\Lambda=\SI{1000}{\GeV}$. Shown are results at \LO{} and \NLO{} in \QCD{} and at $1/\Lambda^2$ in the \SMEFT{} 
		as 
		well as higher order effects $1/\Lambda^4$.}
	\label{fig:cosyprodimo3}
\end{figure}

  While \emph{inclusively}
the operator contribution for $\Imag\,\Cthree = 1$ and
$\Lambda = \SI{1000}{\GeV}$ is tiny at $1/\Lambda^2$, this is not
true \emph{differentially}.  At \LO{} the $1/\Lambda^4$ contributions
are small, but they are not flat, and the \NLO{} contribution is
neither small nor flat.  We display these issues
in \cref{fig:cosxprodimo3}
and \cref{fig:cosyprodimo3}. \Cref{fig:cosxprodimo3} shows with the
$\cosxprod$ distribution that inclusively the contribution from
$\Imag\,\Cthree$ is small at \LO{}, but it is
large at \NLO{} \QCD{} with a complex angular structure.  The $\cosyprod$
distribution in \cref{fig:cosyprodimo3} shows that differentially in
$\cosyprod$ at \NLO{}, $\Imag\,\Cthree$ leads to large
contributions, where large negative contributions for $\cosyprod < 0$
cancel with positive contributions in $\cosyprod > 0$ inclusively, but
will strongly modify the \SM{} result. 
Interestingly the shift in
$\cosyprod$ here is similar in shape and size to the shift seen
in \cref{fig:cosyprodsm} when going from an on-shell approximation to
the full off-shell result in the \SM{}.

\paragraph{The sensitivity to $1/\Lambda^2$ contributing \SMEFT{} operators.}

One way to present the \emph{sensitivity} of the angular observables
to the various operators is to show normalized distributions where
the \SM{} contributions are divided out. As previous we compute
the \SMEFT{} contribution $\sigma^\EFT{}$, but now we also normalize
these distributions to themselves and divide by the normalized \SM{}
distribution.  In this way one can see the shape difference of
the \SMEFT{} contribution with respect to the \SM{}. Since shape
differences in differential distributions result in the highest
discriminating power, operators with stronger shape differentials can
be constrained better.  In short, what we show in the following for a
differential distribution with respect to the variable $x$ is
\begin{equation}
	\left( \frac{1}{\sigma^\EFT{}} \cdot \frac{\Delta\sigma^\EFT{}}{\Delta x} \right) 
	 \biggl/
	\left( \frac{1}{\sigma^\SM{}} \cdot \frac{\Delta\sigma^\SM{}}{\Delta x} \right)\,,
\end{equation}
where $\sigma^\EFT{}$ is the \SM{} subtracted pure \SMEFT{}
contribution, $\sigma = \sigma^\EFT{} + \sigma^\SM{}$.

\begin{figure}
	\centering\includegraphics{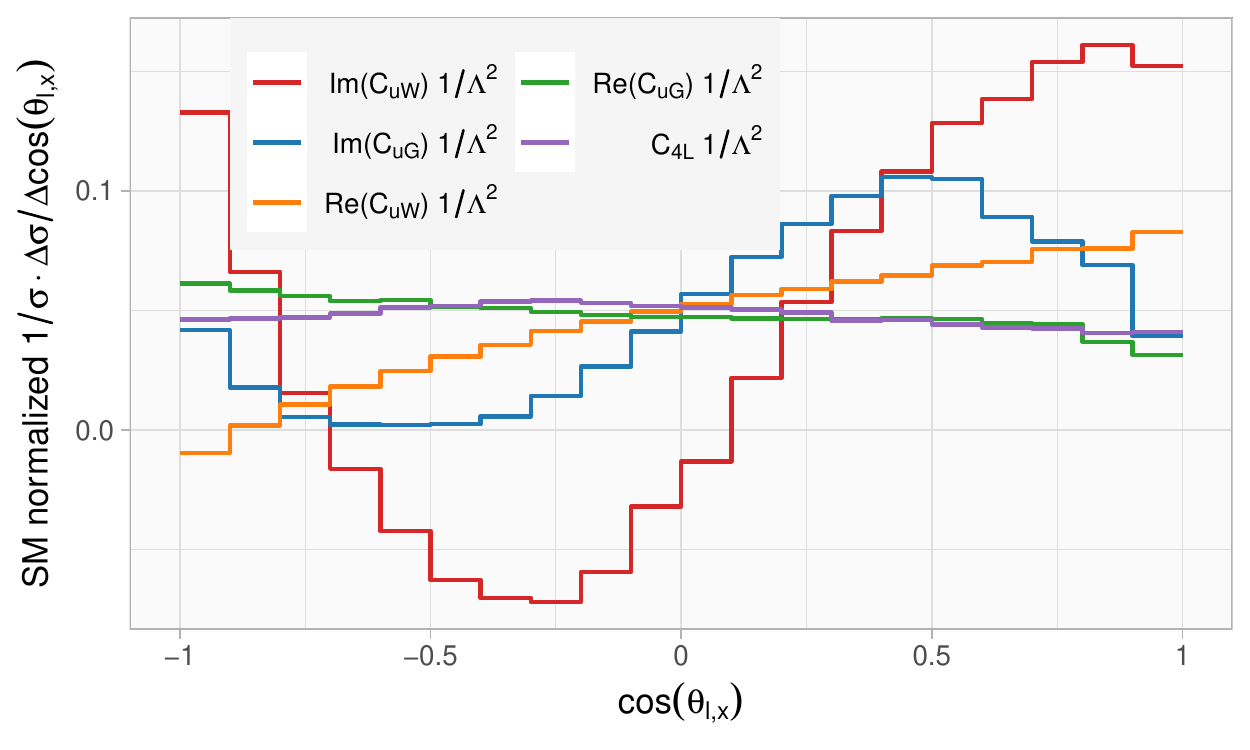}
	\caption{To itself and to the \SM{} normalized \SMEFT{} contributions to the $\cosxprod$ distribution at \NLO{} in 
	\QCD{}. Results are presented for the real and imaginary parts of $\Cthree$ and $\Csix$ as well as for the four 
	quark operator $\Ceight$.}
	\label{fig:cosxprodall}
\end{figure}

In \cref{fig:cosxprodall} we show the resulting shapes for the real
and imaginary parts of the Wilson coefficients for the operators
$\Qthree$ and $\Qsix$ and the four quark operator $\Qeight$. Note that
the operator $\Qsix$ has no \LO{} contribution since it only enters
at \NLO{}.  Generally $\Imag\,\Cthree$ and
$\Imag\,\Csix$ have the largest shape differentials. Arriving
at precise limits on Wilson coefficients is thus strongly influenced
by unique \NLO{} contributions.  The corresponding distributions for
the angles $\cosyprod$, $\coszprod$ and $\coslstar$ as well as
$\coslT$ are in included in \cref{sec:appendix}
in \cref{fig:cosyprodall,fig:coszprodall,fig:coslTprodall,fig:coslstarprodall}
for completeness, and are maximally sensitive to the imaginary parts
of the operators.

\paragraph{The sensitivity to $1/\Lambda^4$ contributing \SMEFT{} operators.}

We treat operators that enter at $1/\Lambda^4$ as leading order in the
\SMEFT{} in the sense that we do not consider possible mixing with
dimension eight operators that appear at \NLO{} in \QCD{}.  We
include these operators to allow for a mapping onto the right-handed
anomalous coupling experimental limits, such as arise in $W^\prime$ models
\cite{Tait:2000sh,Sullivan:2002jt,Duffty:2012rf}.  These operators do not
interfere with the \SM{} amplitude.

We present self-normalized angular distributions that are divided by
the \SM{} angular distribution shape for the operators
$\Qtwo, \Qfour, \Qseven$ and $\Qnine$ in
\cref{fig:cosxprodl4all,fig:cosyprodl4all,fig:coszprodl4all,fig:coslTl4all,fig:coslstar} 
in \cref{sec:appendix}.  The cross sections only depend on the modulus
squared of the Wilson coefficients $|C_i|^2$.  Operators like $\Qtwo$
that act like a right-handed $Wtb$ coupling predictably create a
strong $\coszprod$ dependence that can be distinguished from the
four-fermion operator $\Qnine$ by comparing to $\cosyprod$.


\section{Conclusions}

In this study we present for the first time a full $2\to 4$ off-shell
calculation of $t$-channel single-top-quark production at \NLO{}
in \QCD{}, taking into account the decay of the $W$-boson at the
amplitude level.  We use the complex mass scheme to gauge-invariantly
include such off-shell effects in both the \SM{}, and in the \SMEFT{}
framework. We include all relevant dimension six operators that affect 
the $Wtb$ vertex at \NLO{} in \QCD{}.

We examine off-shell effects in the \SM{} and find significant
differences with respect to the on-shell approximation.  We also
consider effects due to the reconstruction of the $W$-boson and
neutrino coincident with the off-shell effects.  While sensitivity of
the reconstructed top-quark invariant mass distribution near its peak
due to soft radiation is expected for an off-shell top quark; we point
out for the first time how this carries through to certain angular
distributions used by experimental analyses.
In particular, we find that the $\coslN$ distribution used by some
analyses, which is constructed in the top-quark rest frame, becomes
unphysical when off-shell effects are included.  This problem is
hidden in the on-shell calculations.  Without a resummation of soft
radiation in the on-shell region, we cannot recommend the use of this
observable for precision studies.

Our results move beyond the common \LO{} \SMEFT{} picture and allow for
a fully consistent \SMEFT{} evaluation at $1/\Lambda^2$, as well as
partial corrections from $1/\Lambda^4$ operators in order to compare
to the anomalous couplings picture.  We show that \NLO{} \QCD{}
effects to the \SMEFT{} contributions can be large in angular
distributions, and in general are not captured by a rescaling
with \SM{} $K$-factors.  In addition, the operators $\Qsix$ and
$\Qseven$ are included for the first time in the full process
at \NLO{}.  These operators only begin to enter at \NLO{} in \QCD{},
and are important for consistent \NLO{} limits on Wilson coefficients.

We present an extensive list of checks of our amplitudes, and we
successfully compare several \SM{} and \SMEFT{} terms at \LO{}
and \NLO{} with results from the literature. We are able to identify
the correct relative signs of all pieces of our calculation through
checks of gauge-invariance, and \UV{} and \IR{} finiteness.  Because
of the asymmetric nature of limits for positive and negative \SMEFT{}
operators that arise in the \NLO{} calculations, it is imperative for
experimental analyses to clearly state which sign convention is used
for both minimal coupling in the \SM{} and for the \SMEFT{}.

While we present the most complete fixed-order perturbative
calculation of the \SMEFT{} operators, a dedicated study of parton
shower effects to observables used for \SMEFT{} studies would be
useful.  Parton shower effects are included in
ref.~\cite{deBeurs:2018pvs}, but they do not discuss their impact on
the fixed order results.  Some observables, like the top-quark
invariant mass distribution clearly need to include all-order effects.
Our study is also performed with a massless bottom quark, but bottom-quark
mass effects could become interesting at future precision levels.

Our implementation is publicly available in \MCFM{}-8.3 and includes
preconfigured plotting routines to reproduce all distributions in this
study.  It is our goal to allow for easy, yet precise and refined
determination of observables in the \SM{} and better constraints
on \SMEFT{} operators.

\paragraph{Acknowledgments.}

We thank Stefan Dittmaier for helpful comments on the reduction to
spinor chain master structures, John Campbell for invaluable
continuous discussion, Bogdan Dobrescu for useful discussion on model
interpretation of the \EFT{} operators, Claudius Krause for helpful
discussion about the \SMEFT{}, and Mario Prausa for help on
interpreting Fermat output. We would also like to thank Cen Zhang
for helping to track down the sign discrepancy we stated in the preprint of
this study to a different sign convention used in parts of the literature.
Feynman diagrams were generated with TikZ-Feynman \cite{Ellis:2016jkw}.

This work was supported by the U.S.\ Department of Energy under award
No.\ DE-SC0008347.  This document was prepared using the resources of
the Fermi National Accelerator Laboratory (Fermilab), a
U.S. Department of Energy, Office of Science, HEP User
Facility. Fermilab is managed by Fermi Research Alliance, LLC (FRA),
acting under Contract No.\ DE-AC02-07CH11359.

\newpage

\appendix

\section{Additional figures}
\label{sec:appendix}

\begin{figure}
	\centering\includegraphics{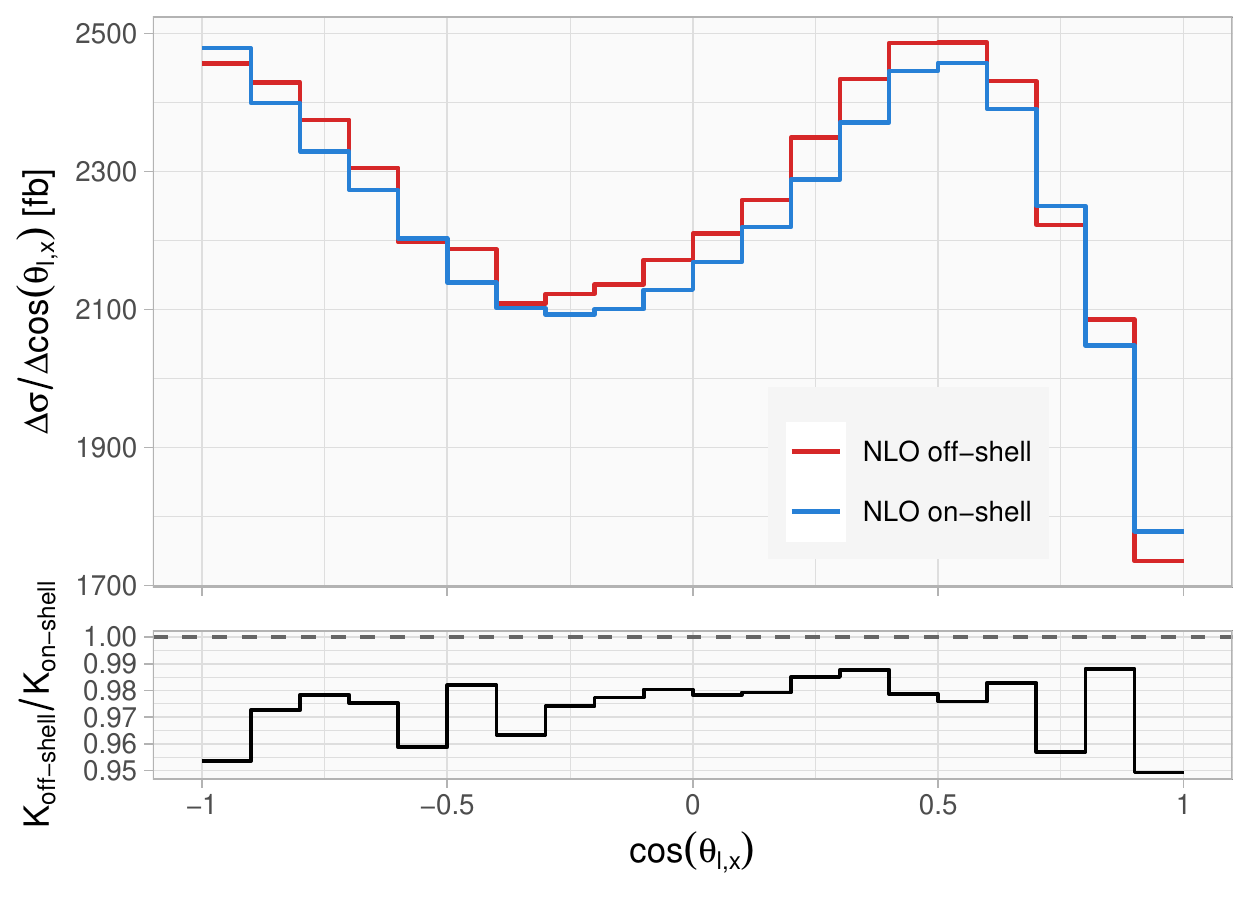}
	\caption{Distribution of $\cosxprod$ at \NLO{} in the on-shell approximation and for an off-shell top quark. The 
	lower panel shows the ratio of the off-shell and on-shell $K$-factors. Note the range of the $y$-axis, which does 
	not start at zero.}
	\label{fig:cosxprodsm}
\end{figure}

\begin{figure}
	\centering\includegraphics{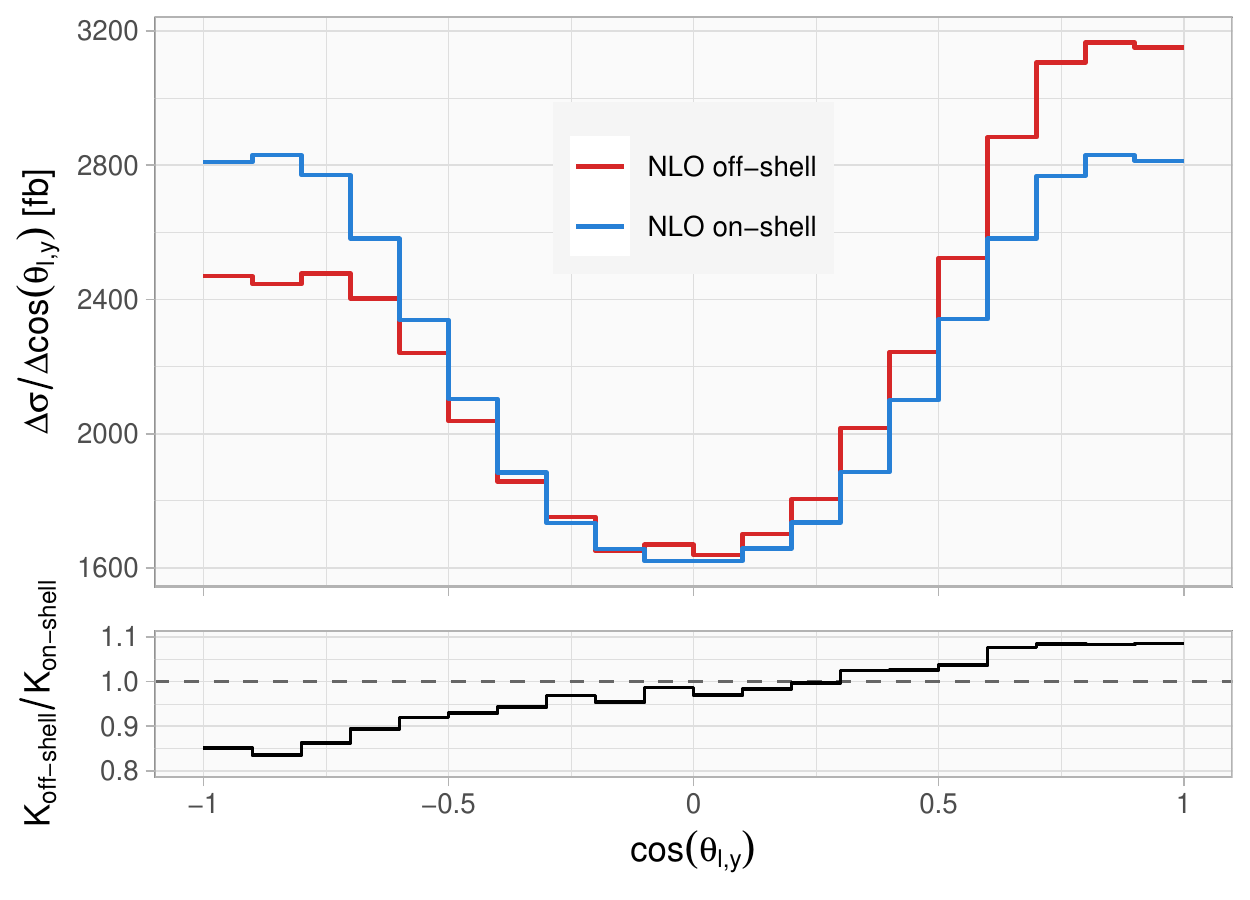}
	\caption{Distribution of $\cosyprod$ at \NLO{} in the on-shell approximation and for an off-shell top quark. The 
		lower panel shows the ratio of the off-shell and on-shell $K$-factors. Note the range of the $y$-axis, which 
		does not start at zero.}
	\label{fig:cosyprodsm}
\end{figure}

\begin{figure}
	\centering\includegraphics{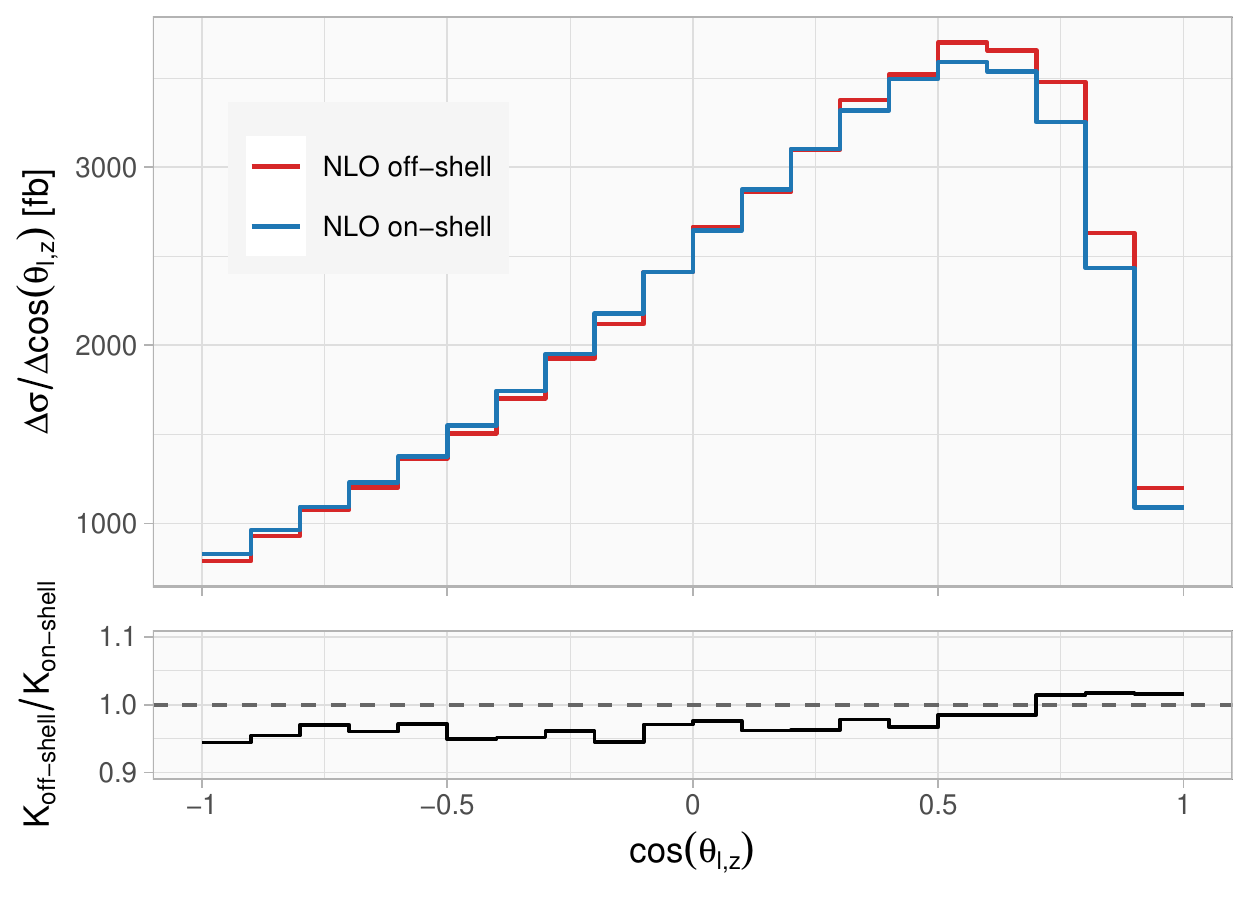}
	\caption{Distribution of $\coszprod$ at \NLO{} in the on-shell approximation and for an off-shell top quark. The 
		lower panel shows the ratio of the off-shell and on-shell $K$-factors. Note the range of the $y$-axis, which 
		does not start at zero.}
	\label{fig:coszprodsm}
\end{figure}

\begin{figure}
	\centering\includegraphics{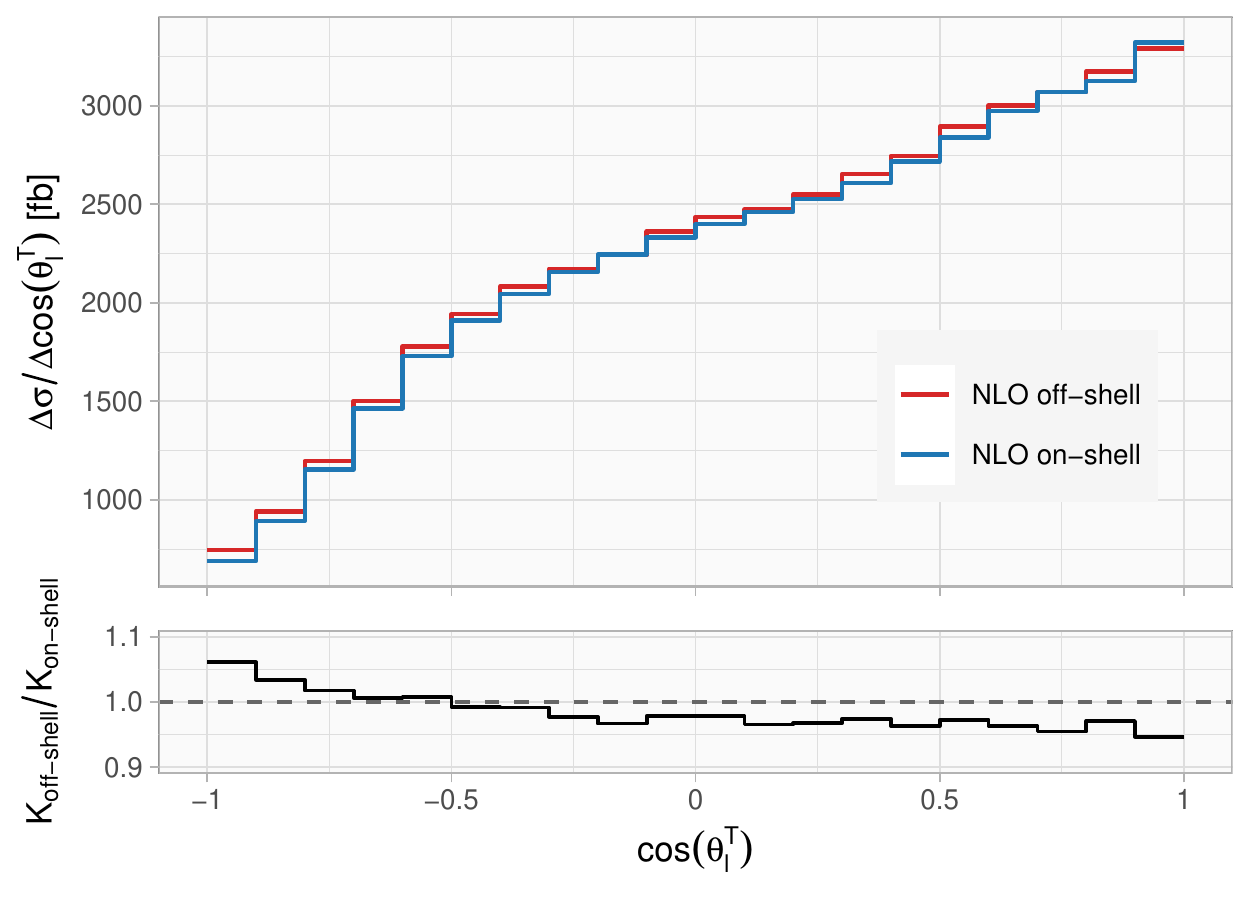}
	\caption{Distribution of $\coslT$ at \NLO{} in the on-shell approximation and for an off-shell top quark. The 
		lower panel shows the ratio of the off-shell and on-shell $K$-factors. Note the range of the $y$-axis, which 
		does not start at zero.}
	\label{fig:coslTsm}
\end{figure}

\begin{figure}
	\centering\includegraphics{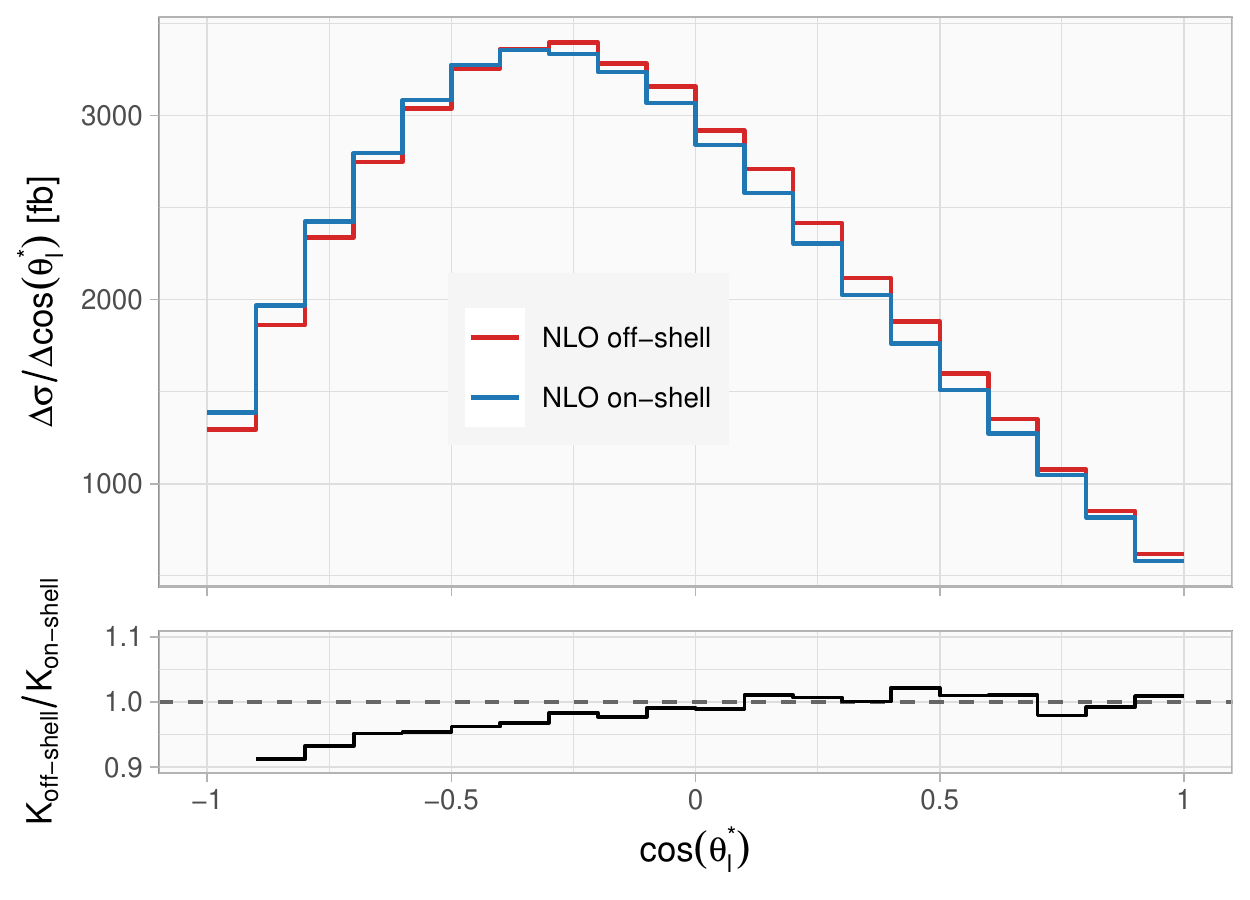}
	\caption{Distribution of $\coslstar$ at \NLO{} in the on-shell approximation and for an off-shell top quark. The 
		lower panel shows the ratio of the off-shell and on-shell $K$-factors. Note the range of the $y$-axis, which 
		does not start at zero.}
	\label{fig:coslstarsm}
\end{figure}


\begin{figure}
	\centering\includegraphics{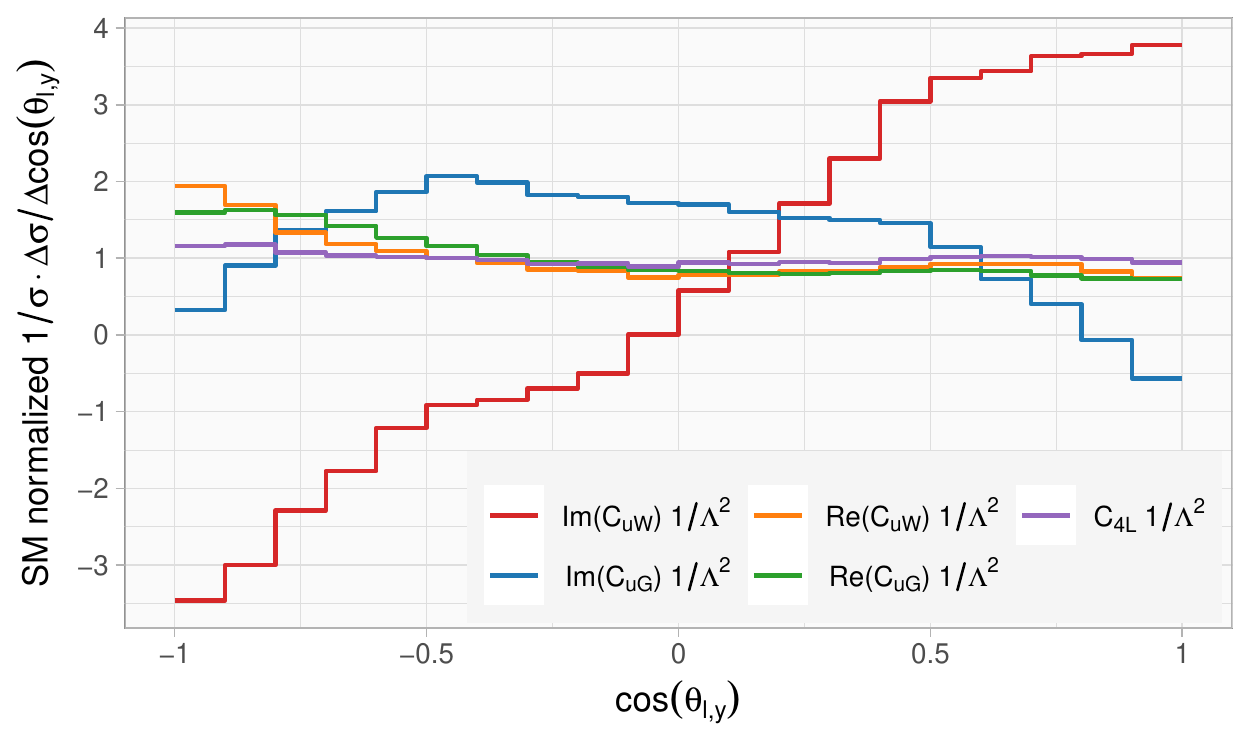}
	\caption{To itself and to the \SM{} normalized \SMEFT{} contributions to the $\cosxprod$ distribution at \NLO{} in 
		\QCD{}. Results are presented for the real and imaginary parts of $\Cthree$ and $\Csix$ as well as for the four 
		quark operator $\Ceight$.}
	\label{fig:cosyprodall}
\end{figure}

\begin{figure}
	\centering\includegraphics{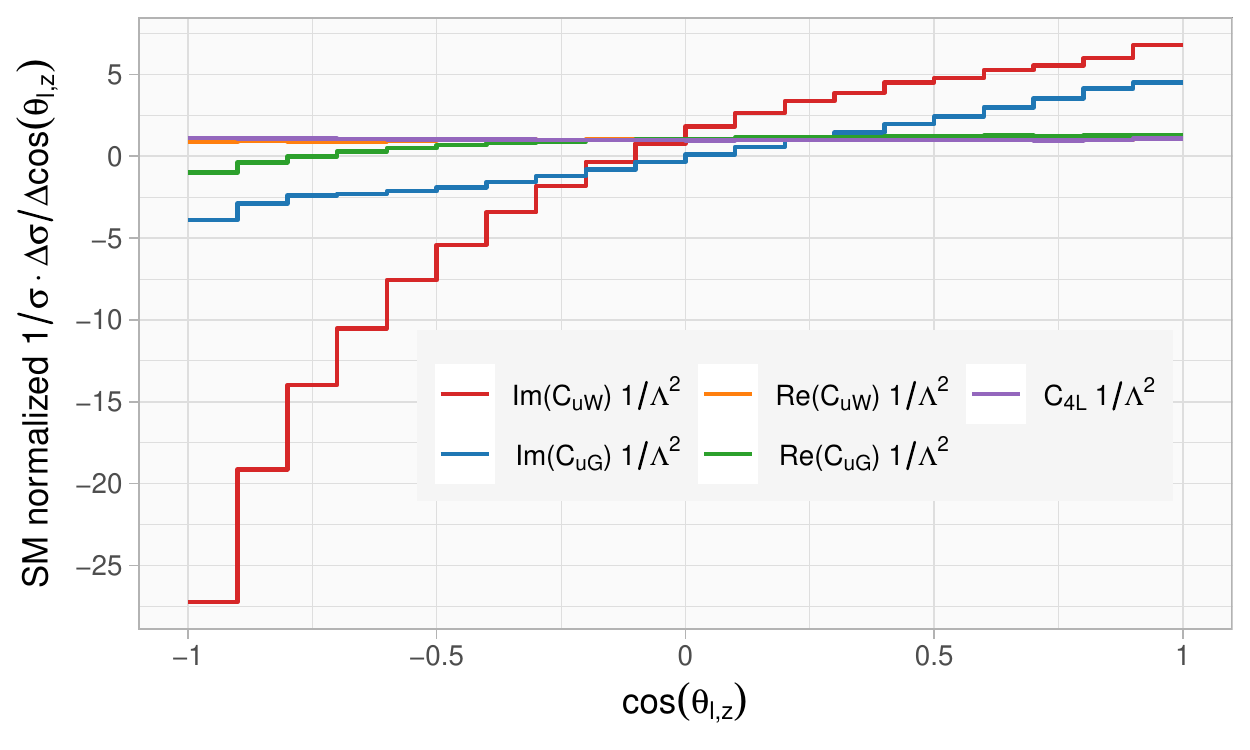}
	\caption{To itself and to the \SM{} normalized \SMEFT{} contributions to the $\cosyprod$ distribution at \NLO{} in 
		\QCD{}. Results are presented for the real and imaginary parts of $\Cthree$ and $\Csix$ as well as for the four 
		quark operator $\Ceight$.}
	\label{fig:coszprodall}
\end{figure}

\begin{figure}
	\centering\includegraphics{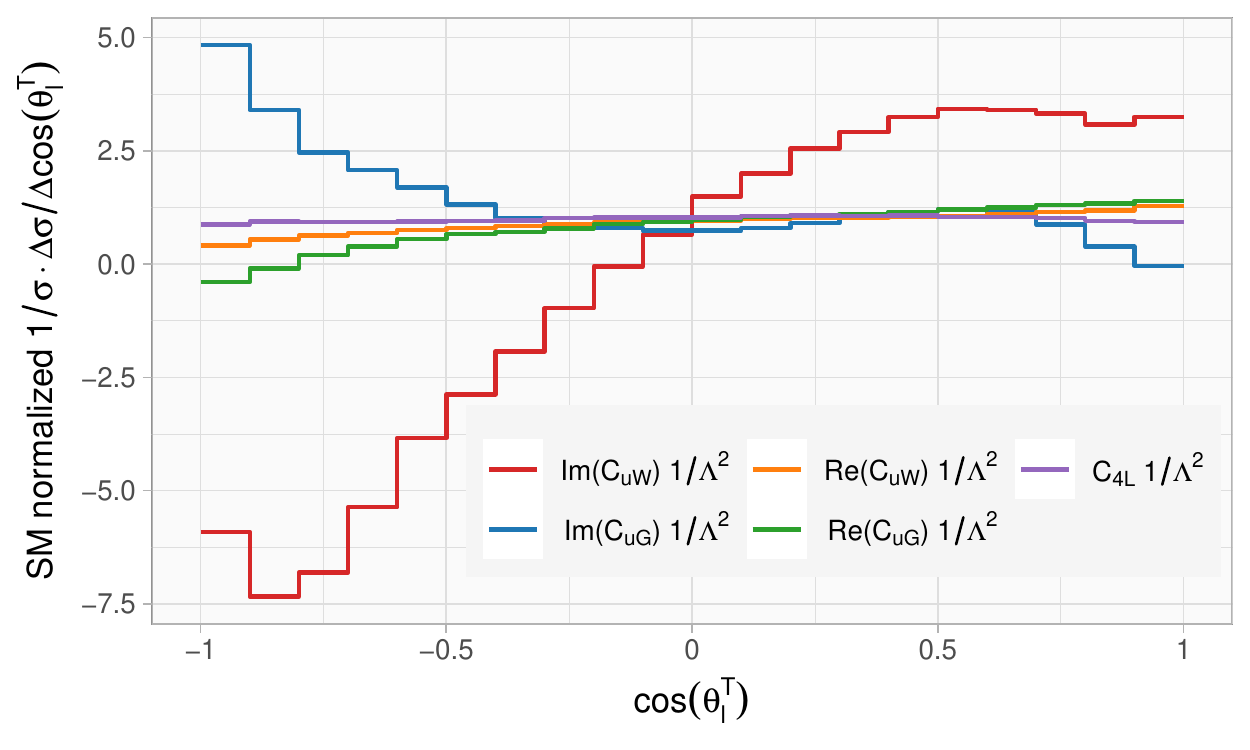}
	\caption{To itself and to the \SM{} normalized \SMEFT{} contributions to the $\coslT$ distribution at \NLO{} in 
		\QCD{}. Results are presented for the real and imaginary parts of $\Cthree$ and $\Csix$ as well as for the four 
		quark operator $\Ceight$.}
	\label{fig:coslTprodall}
\end{figure}

\begin{figure}
	\centering\includegraphics{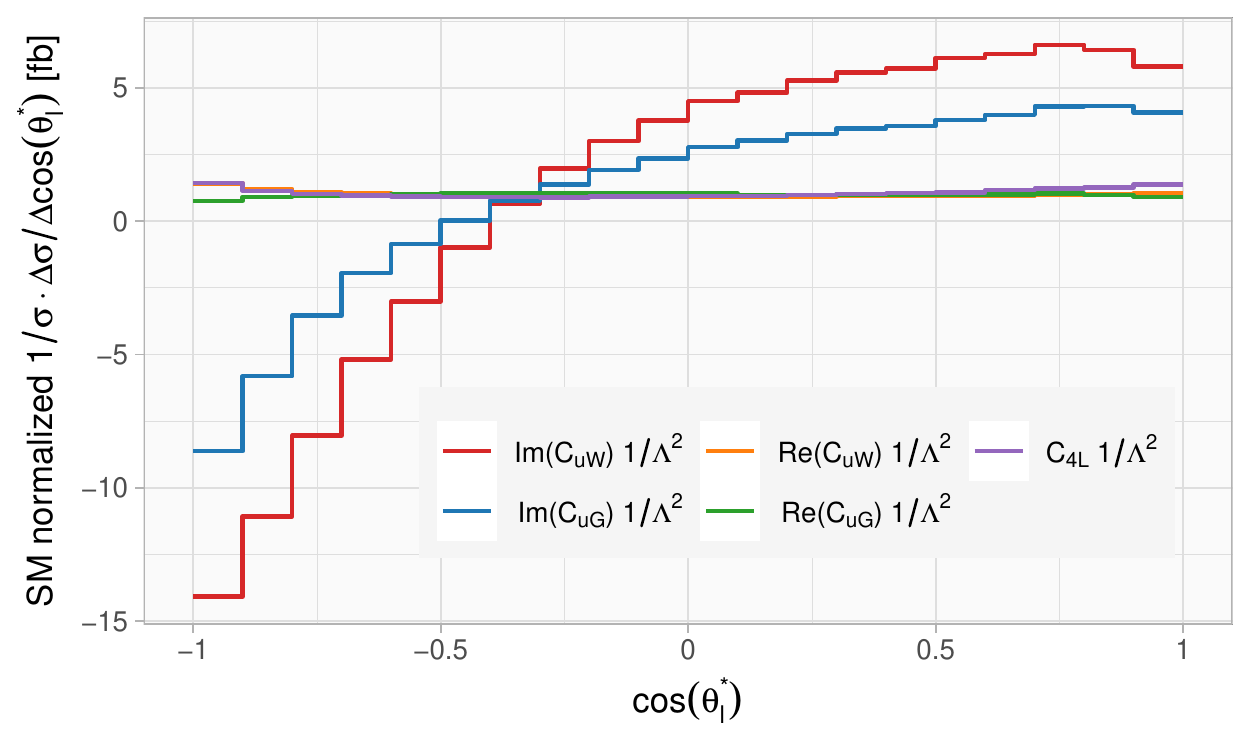}
	\caption{To itself and to the \SM{} normalized \SMEFT{} contributions to the $\coslstar$ distribution at \NLO{} in 
		\QCD{}. Results are presented for the real and imaginary parts of $\Cthree$ and $\Csix$ as well as for the four 
		quark operator $\Ceight$.}
	\label{fig:coslstarprodall}
\end{figure}


\begin{figure}
	\centering\includegraphics{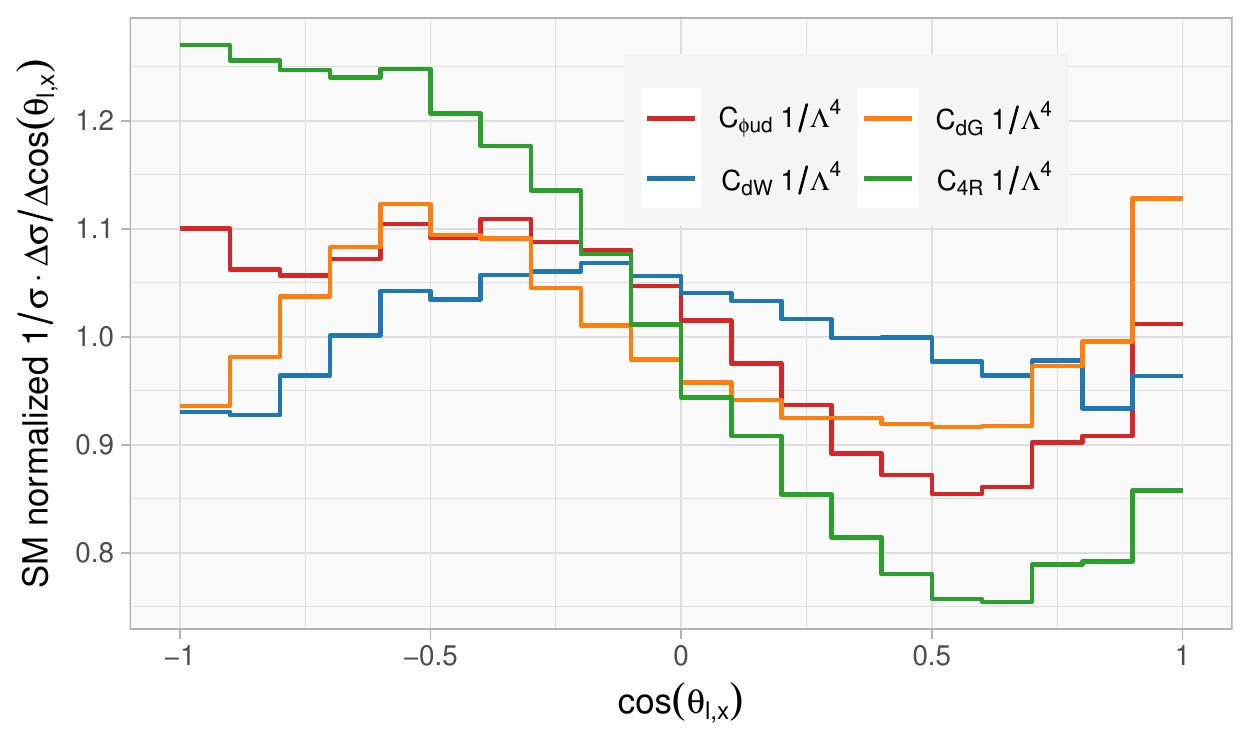}
	\caption{To itself and to the \SM{} normalized \SMEFT{} contributions to the $\cosxprod$ distribution at \NLO{} in 
		\QCD{}. Results are presented for $\Qtwo$, $\Qfour$ and $\Qseven$ as well as for the four quark operator 
		$\Qnine$.}
	\label{fig:cosxprodl4all}
\end{figure}

\begin{figure}
	\centering\includegraphics{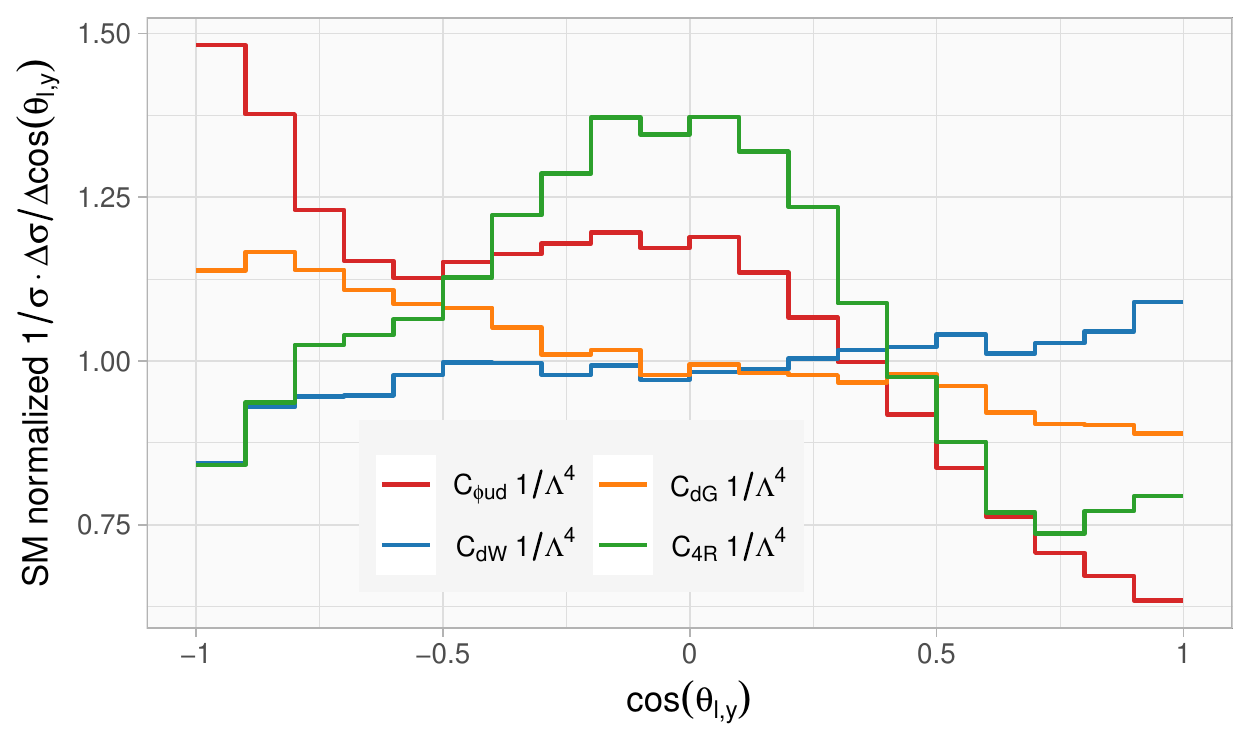}
	\caption{To itself and to the \SM{} normalized \SMEFT{} contributions to the $\cosyprod$ distribution at \NLO{} in 
		\QCD{}. Results are presented for $\Qtwo$, $\Qfour$ and $\Qseven$ as well as for the four quark operator 
		$\Qnine$.}
	\label{fig:cosyprodl4all}
\end{figure}

\begin{figure}
	\centering\includegraphics{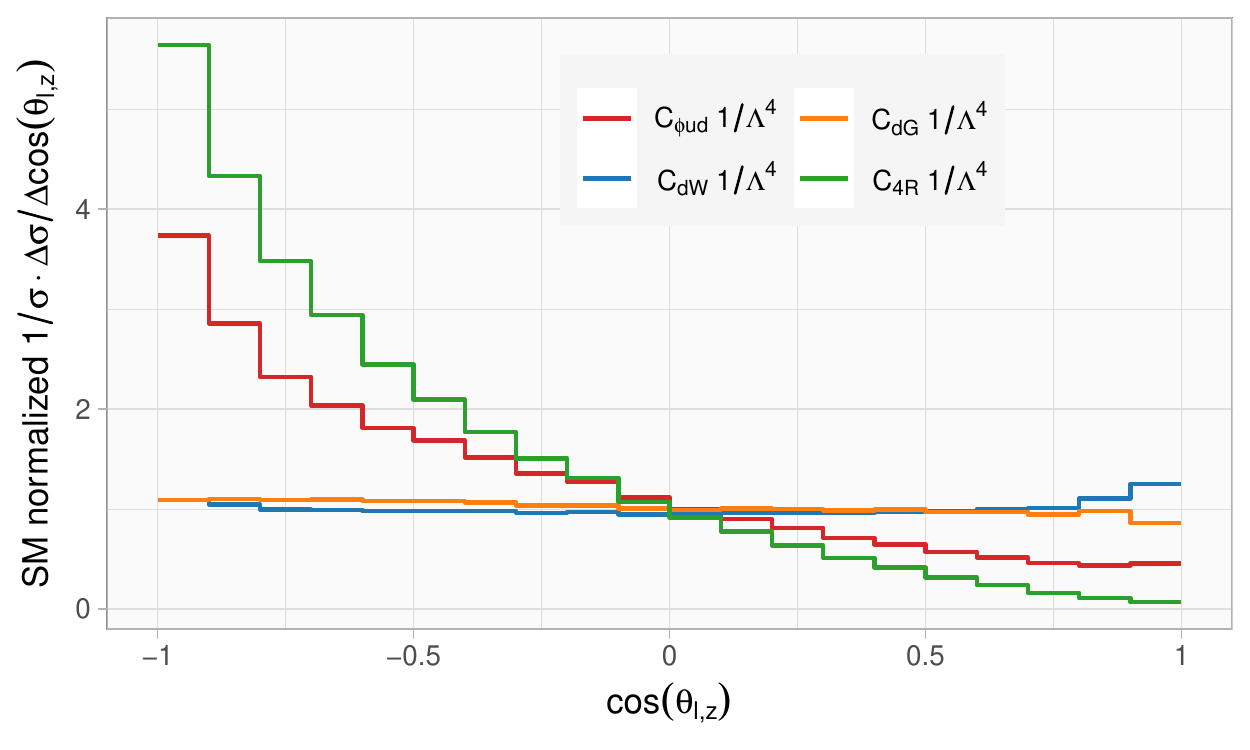}
	\caption{To itself and to the \SM{} normalized \SMEFT{} contributions to the $\coszprod$ distribution at \NLO{} in 
		\QCD{}. Results are presented for $\Qtwo$, $\Qfour$ and $\Qseven$ as well as for the four quark operator 
		$\Qnine$.}
	\label{fig:coszprodl4all}
\end{figure}

\begin{figure}
	\centering\includegraphics{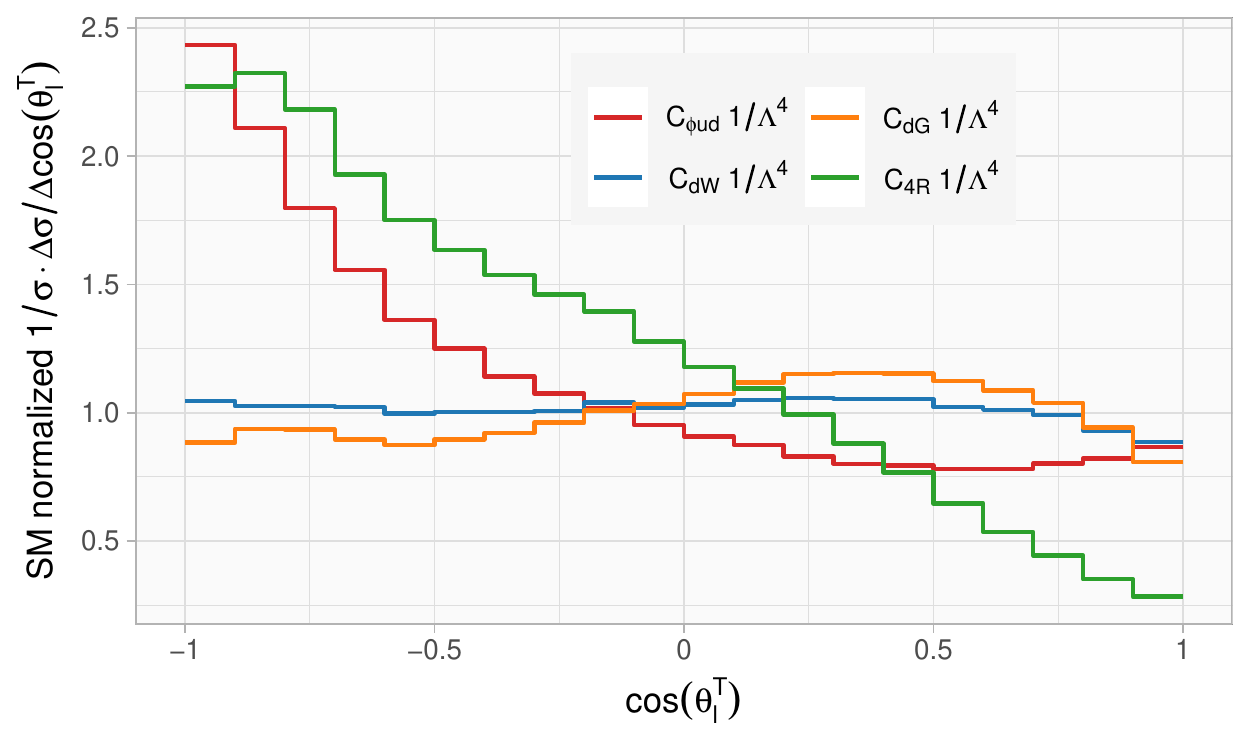}
	\caption{To itself and to the \SM{} normalized \SMEFT{} contributions to the $\coslT$ distribution at \NLO{} in 
		\QCD{}. Results are presented for $\Qtwo$, $\Qfour$ and $\Qseven$ as well as for the four quark operator 
		$\Qnine$.}
	\label{fig:coslTl4all}
\end{figure}

\begin{figure}
	\centering\includegraphics{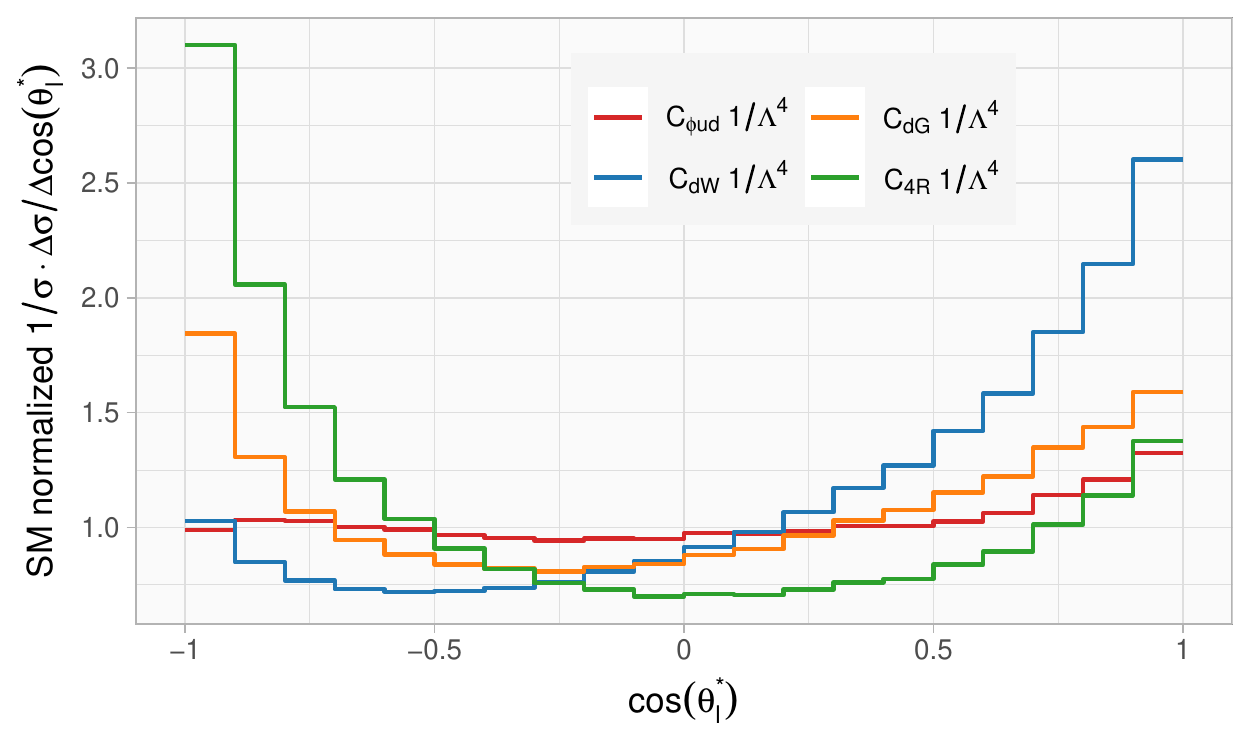}
	\caption{To itself and to the \SM{} normalized \SMEFT{} contributions to the $\coslstar$ distribution at \NLO{} in 
		\QCD{}. Results are presented for $\Qtwo$, $\Qfour$ and $\Qseven$ as well as for the four quark operator 
		$\Qnine$.}
	\label{fig:coslstar}
\end{figure}

\newpage

\bibliographystyle{JHEP}
\bibliography{bib}

\end{document}